\documentclass[modern]{aastex63}
\usepackage{lineno}

\usepackage{url}
\usepackage[section]{placeins}
\usepackage{longtable}
\usepackage{rotating}
\usepackage{pgfplotstable}
\newcommand{\epc}{$\epsilon$ Cnc}
\newcommand{\kms}{km s$^{-1}\;$}
\newcommand{\msun}{M_\odot}

\newcommand{\Lsun}{L_\odot}
\newcommand{\teff}{T_{\rm eff}}

\begin{document}

\title{The Interferometric Binary $\epsilon$ Cnc in Praesepe: Precise Masses and Age}

\author[0000-0002-1333-8866]{Leslie M. Morales}
\affiliation{San Diego State University, Department of Astronomy, San
  Diego, CA, 92182 USA}
  
\author[0000-0003-4070-4881]{Eric L. Sandquist}
\affiliation{San Diego State University, Department of Astronomy, San
  Diego, CA, 92182 USA}

\author[0000-0001-5415-9189]{Gail H. Schaefer}
\affiliation{The CHARA Array of Georgia State University, Mount Wilson Observatory, Mount Wilson, CA 91023, USA}

\author[0000-0001-9939-2830]{Christopher D. Farrington}
\affiliation{The CHARA Array of Georgia State University, Mount Wilson Observatory, Mount Wilson, CA 91023, USA}

\author[0000-0002-4313-0169]{Robert Klement}
\affiliation{The CHARA Array of Georgia State University, Mount Wilson Observatory, Mount Wilson, CA 91023, USA}

\author[0000-0003-4080-6466]{Luigi R. Bedin}
\affiliation{Istituto Nazionale Astrofisica di Padova—Osservatorio Astronomico di Padova, Vicolo dell'Osservatorio 5, I-35122 Padova, Italy}

\author[0000-0001-9673-7397]{Mattia Libralato}
\affiliation{AURA for the European Space Agency (ESA), ESA Office, Space Telescope Science Institute, 3700 San Martin Drive, Baltimore, MD 21218, USA}

\author[0000-0002-6492-2085]{Luca Malavolta}
\affiliation{Istituto Nazionale Astrofisica di Padova—Osservatorio Astronomico di Padova, Vicolo dell'Osservatorio 5, I-35122 Padova, Italy}
\affiliation{Dipartimento di Fisica e Astronomia "Galileo Galilei",Università di Padova, Vicolo dell'Osservatorio 3, Padova I-35122, Italy}

\author[0000-0003-1149-3659]{Domenico Nardiello}
\affiliation{Istituto Nazionale Astrofisica di Padova—Osservatorio Astronomico di Padova, Vicolo dell'Osservatorio 5, I-35122 Padova, Italy}
\affiliation{Aix Marseille Univ, CNRS, CNES, LAM, Marseille, France}

\author[0000-0001-9647-2886]{Jerome A. Orosz}
\affiliation{San Diego State University, Department of Astronomy, San
  Diego, CA, 92182 USA}
  
\author[0000-0002-3380-3307]{John D. Monnier}
\affiliation{Astronomy Department, University of Michigan, Ann Arbor, MI 48109, USA}

\author[0000-0001-6017-8773]{Stefan Kraus}
\affiliation{Astrophysics Group, Department of Physics \& Astronomy, University of Exeter, Stocker Road, Exeter, EX4 4QL, UK}

\author[0000-0002-0493-4674]{Jean-Baptiste Le Bouquin}
\affiliation{Institut de Planetologie et d'Astrophysique de Grenoble, Grenoble 38058, France}
\affiliation{Astronomy Department, University of Michigan, Ann Arbor, MI 48109, USA}

\author[0000-0002-2208-6541]{Narsireddy Anugu}
\affiliation{The CHARA Array of Georgia State University, Mount Wilson Observatory, Mount Wilson, CA 91023, USA}

\author{Theo ten Brummelaar}
\affiliation{The CHARA Array of Georgia State University, Mount Wilson Observatory, Mount Wilson, CA 91023, USA}

\author[0000-0001-9764-2357]{Claire L. Davies}
\affiliation{Astrophysics Group, Department of Physics \& Astronomy, University of Exeter, Stocker Road, Exeter, EX4 4QL, UK}

\author[0000-0002-1575-4310]{Jacob Ennis}
\affiliation{Astronomy Department, University of Michigan, Ann Arbor, MI 48109, USA}

\author[0000-0002-3003-3183]{Tyler Gardner}
\affiliation{Astronomy Department, University of Michigan, Ann Arbor, MI 48109, USA}

\author[0000-0001-9745-5834]{Cyprien Lanthermann}
\affiliation{The CHARA Array of Georgia State University, Mount Wilson Observatory, Mount Wilson, CA 91023, USA}

\begin{abstract}
We observe the brightest member of the Praesepe cluster, \epc, to precisely measure the characteristics of the stars in this binary system, en route to a new measurement of the cluster's age. We present spectroscopic radial velocity measurements and interferometric observations of the sky-projected orbit to derive the masses, which we find to be 
$M_1/M_{\odot} = 2.420 \pm 0.008$ and $M_2/M_{\odot} = 2.226 \pm 0.004$. We place limits on the color-magnitude positions of the stars by using spectroscopic and interferometric luminosity ratios while trying to reproduce the spectral energy distribution of \epc.
We re-examine the cluster membership of stars at the bright end of the color-magnitude diagram using Gaia data and literature radial velocity information. The binary star data are consistent with an age of  $637\pm19$ Myr, as determined from MIST model isochrones. The masses and luminosities of the stars appear to select models with the most commonly used amount of convective core overshooting.
 
\end{abstract}

\section{Introduction} \label{sec:intro}

Absolute age measurements for stars are fraught with difficulties because comparisons with theoretical models are required, and because our knowledge of important factors like the initial chemical composition, or the stellar masses, or even the behavior of certain basic physical processes is often poor. Often, even when we are able to precisely measure characteristics of stars that can speak to a star's age, these gaps in our understanding result in substantial systematic errors. Star clusters provide an environment where difficulties like chemical composition uncertainties can be mitigated and a uniform age is usually a good assumption. The study of binary systems within such clusters can give the opportunity to evaluate stellar masses in a precise way.

The goals of this paper are to precisely characterize stars (including masses) at the turnoff of the open cluster Praesepe, and derive a precise age for the cluster from these stars in order to set up Praesepe as an age-calibration standard. There are a number of astrophysically-interesting features of the cluster that recommend it for attention. The rotation of the main sequence stars has been studied extensively, among the A-type stars \citep{mcgee,aandw,debernardi,2007AA...476..911F,fossati08,cummings}, solar-type stars \citep{delorme,kovacs}, and low-mass stars \citep{scholz,agueros,scholz11,douglas17}. These measurements establish Praesepe as a gyrochronology age standard, although a more precise absolute age would improve its utility.
Among the A stars, there are a number of interesting phenomena like $\delta$ Scuti pulsation \citep[e.g.][]{breger,frandsen,hernandez,belmonte} and surface chemical peculiarities \citep{abt86,hui,bc98,2007AA...476..911F,fossati08}. The cluster has a fairly large population of white dwarfs, with 12 known \citep{salaris}. Almost all of the white dwarfs have spectroscopically-measured masses, and they cover a fairly large range from about 0.7 to $0.9 \msun$ \citep{cummingswd,salaris}, in contrast to the field white dwarf population. These white dwarfs probably originate from progenitor stars in a fairly narrow mass range, where the stars ignite core helium burning in a non-degenerate gas but do not undergo a second dredge-up convection event that helps limit the size of the core.

To these ends, we have conducted a thorough analysis of the known double-lined spectroscopic binary star system \epc \  (also known as Meleph, HD 73731, 41 Cnc, BD +20 2171). With the combined light of its stars, \epc \ is one of the brightest members of Praesepe. \citet{plaskett} were the first to measure the velocity of the second component of the binary, but \citet{aandw} were the first to fit the spectroscopic orbit of the system, finding a period of 35.202 days. Both stars in the binary are rotating relatively modestly for A-type stars, and both stars appear to show surface chemical peculiarities \citep{bc98}. But most importantly, the binary is bright enough, close enough, and has a wide enough orbit that the orbit can now be resolved with interferometric methods and a precise measurement of the masses of the component stars is within reach. Further, the system's color implies that both stars are on or near the very brightest part of the main sequence, with the brighter star likely to be near the end of its core hydrogen burning life. Because the brighter star is starting to evolve more rapidly than its companion, its characteristics are sensitive functions of age. 

In section \ref{sec:data}, we discuss the spectroscopic, photometric, and interferometric data used, as well as the measurement of the radial velocities of the stars. In section \ref{sec:process}, we describe the fitting method used to determine the characteristics of the stars, like the masses. In section \ref{sec:disc}, we discuss the age measurement for Praesepe and its applications to other areas of stellar astrophysics, like the white dwarf initial-final mass relation.
 
\subsection{Reddening, Distance, and Metallicity}\label{reddist}

\citet{taylor} critically examined measurements of Praesepe's
reddening, and settled on $E(B-V) = 0.027\pm0.004$. This is consistent with the reddening derived by \citet{gaiahrs} from isochrone fitting. Zero reddening is sometimes used for the cluster \citep[e.g.][]{cummings,cummingswd}, but we believe this is ruled out by comparisons with the Hyades cluster. The chemical compositions and ages of these two clusters are extremely similar, but without a significant reddening for Praesepe, the main sequences do not align, and there are troubling offsets between the white dwarf initial-final mass sequences \citep{salaris}. \citet{douglas19} used a spectroscopic examination of FGK stars in Praesepe and Hyades, and found a reddening value $E(B-V)=0.011\pm0.004$ between zero and the Taylor value.

The weighted average parallax for Praesepe members in {\it Gaia} DR2
was determined as $5.371\pm0.003$ mas \citep{gaiacmd}, or
$5.361\pm0.005$ mas \citep{cg18}.  There are a number of studies
indicating that {\it Gaia} parallaxes are systematically offset to
smaller values (and implying that objects are systematically closer
than {\it Gaia} indicates; \citealt{lindegren,zinn}), however. In this paper, we employ the {\it Gaia} Early Data Release 3 (EDR3), which may have a smaller parallax offset of $-17 \mu$as than DR2 \citep{lindegren3}. Based on a simple selection of members 
within $5\degr$ of the cluster center, 6 mas yr$^{-1}$ of the cluster's proper motion vector, and parallax within 0.4 mas of an iteratively-determined average, we find $\bar{\omega} = 5.424$ mas.
This implies a distance modulus $(m-M)_0 = 6.33\pm0.02$.

Recent spectroscopic examinations of Praesepe stars agree well that the cluster is quite metal-rich, similar to the Hyades. Most of the narrow-lined A-type stars in the cluster show chemical peculiarities (e.g. \citealt{bc98}), and so are not appropriate for indications of the mean cluster abundances. However, giants and solar-type stars should have deep enough surface convection zones that chemical diffusion effects would be mostly erased. \citet{yang15} find [Fe/H]$=+0.16\pm0.06$ from four giant stars, while \citet{cp11} found [Fe/H]$=+0.16\pm0.05$ from three giants. \citet{boes13} find [Fe/H]$=+0.12\pm0.04$ from 11 solar-type stars, while \citet{dorazi} find [Fe/H]$=+0.21\pm0.01$ from 10 solar-type dwarfs. Finally, \citet{cummings} find [Fe/H]$=+0.156\pm0.066$ from 39 F, G, and K dwarfs, and \citet{vejar} find [Fe/H]$=+0.21\pm0.02$ from 6 stars with spectral types between G5 and F8.

\subsection{Review of Literature Ages}

In the most recent studies of Praesepe, the age of the cluster has been calculated to range from 590 Myrs to 800 Myrs. This is due to debates over the cluster's metallicity and the effects of stellar rotation. \citet{2009} tested different metallicities in isochrone modeling using PADOVA isochrones,  and derived an age of $757\pm36$ Myr using [Fe/H]$ = +0.2$.

Praesepe and Hyades are often studied together because it is widely understood that these are nearly coeval clusters and because of their similarity in characteristics such as metallicity. \citet{2015} asserted the importance of understanding stellar rotation in discussing these clusters because it can cause chemical mixing in their brightest main sequence stars. They argue that if stellar rotation is neglected, it creates inconsistencies in age in the Bayesian color-magnitude technique that was used to date the clusters using PARSEC models. The age of the cluster becomes more consistent in the models when rotation is taken into account. They derived an older age of $800$ Myr using [Fe/H]$ = +0.12$.  

On the other hand, \citet{2018} used MESA models to examine rotating and non-rotating color-magnitude diagrams of Praesepe. In contrast to \citet{2015}, they conclude that the best-fit isochrone is not one with rotation but rather a younger non-rotating one, estimating an age of $590$ Myr using [Fe/H]$ = +0.24$. 

The relative age difference between Praesepe and the Hyades can also be estimated using gyrochronology of solar-type stars. Hyades stars rotate slower on average than Praesepe stars do, leading to a determination that Prasesepe is about 57 Myr younger than the Hyades \citep{douglas19}.

\section{Observations and Analysis} \label{sec:data}

\subsection{Spectroscopy}\label{spectra}

The majority of the radial velocity measurements in this study are based on a reanalysis of spectra taken as part of a survey of bright Praesepe stars by \citet{aandw}. The spectra were obtained on the Kitt Peak National Observatory (KPNO) 0.9m coud\'{e} feed telescope and spectrograph between 1991 and 1996. The spectra generally covered the wavelength range $436-474.2$ nm, at a resolution of about 0.022 nm. \citeauthor{aandw} obtained a double-lined spectroscopic orbit of \epc \ at the time, although they noted that the rotationally-broadened lines of the two stars are always blended. To improve the accuracy of the radial velocities, we performed a new analysis of the spectra,  as detailed in section \ref{sec:process}.

In addition, we extracted two spectra from the archive of the ELODIE spectrograph at the Observatoire de Haute-Provence (OHP). The spectra ($R=42000$) were taken in 2004 by observer R. Monier, covering the range $385-681$ nm.

Finally, we obtained four new spectra in 2020 with the Fibre-fed Echelle Spectrograph (FIES) on the 2.56m Nordic Optical Telescope (NOT). Two spectra each were obtained near predicted radial velocity extremes in order to better resolve the stars individually, and improve the measurement of the velocity amplitude for later determination of the masses.  We used the medium-resolution ($R=46000$) fiber in our observations, covering the spectral range $370-830$ nm \citep{2014AN....335...41T}, although we only used $440 - 660$ nm in our measurements. We processed the images and extracted one-dimensional spectra using FIESTool (version 1.5.1). The wavelength calibration was done using exposures of the ThAr lamp taken immediately before the science exposures. Individual spectral orders were continuum normalized before being merged.

\subsubsection{Literature and Newly-Measured Radial Velocities}

Literature radial velocities are available for the primary star in \epc \ from several sources, but many of these were rejected due to likely effects of blending of the lines of the two stars. \citet{plaskett} report measurements at four epochs in 1919 and 1920 from the 1.83m Plaskett Telescope at Dominion Astrophysical Observatory (DAO), including one claiming detection of both stars. The detection of double lines occurred approximately at one velocity extremum ($\phi \approx 0.14$), and the velocity separation of the measurements agrees well with our modern solution if a $-6.5$ \kms offset is used. We therefore regard the detection as legitimate, and utilize the measurement with the offset applied. The remaining measurements were discarded as affected by blending.

\citet{abt} reported eight measurements taken at Mt. Wilson Observatory in 1923, although we only used five of these. The remaining three measurements with phases $0.1 < \phi < 0.3$ show clear signs of blending. Finally, there were 10 epochs observed by \citet{aandw} for which the original spectra have not been recovered. In those cases, the reported velocities were still used in the fitting of a velocity solution. These literature values are reported along with our measurements in Tables \ref{tab:1} and \ref{tab2}. Julian dates reported by \citet{plaskett} and UT dates reported by \citet{abt} were converted to heliocentric Julian dates.

Radial velocity measurements were made using broadening functions (BFs; \citealt{rucinski}) derived from the spectra. We used the program BF-rvplotter \citep{Rawls} to calculate the broadening functions for each spectrum after modifying the code to fit rotational broadening functions (rather than Gaussians) to the two star profiles. The BFs are calculated through a comparison with a non-rotating synthetic spectrum of similar temperature and surface gravity, which in this case came from \citet{coelho}. Because of the particular combination of radial velocity separations and rotational broadening for the two stars, we found that the broadening profiles of the stars overlapped substantially and produced significant correlations between the measured rotation and radial velocities. As a result, we found it necessary to determine rotational velocities for the two stars from spectra with the highest signal-to-noise ratios and velocity separations. We subsequently fixed the values of the rotational velocities and refitted all spectra to derive the radial velocities we present. Heliocentric velocity corrections were applied to all calculated velocity values.

\begin{figure}
    \centering
    \includegraphics[scale=.45]{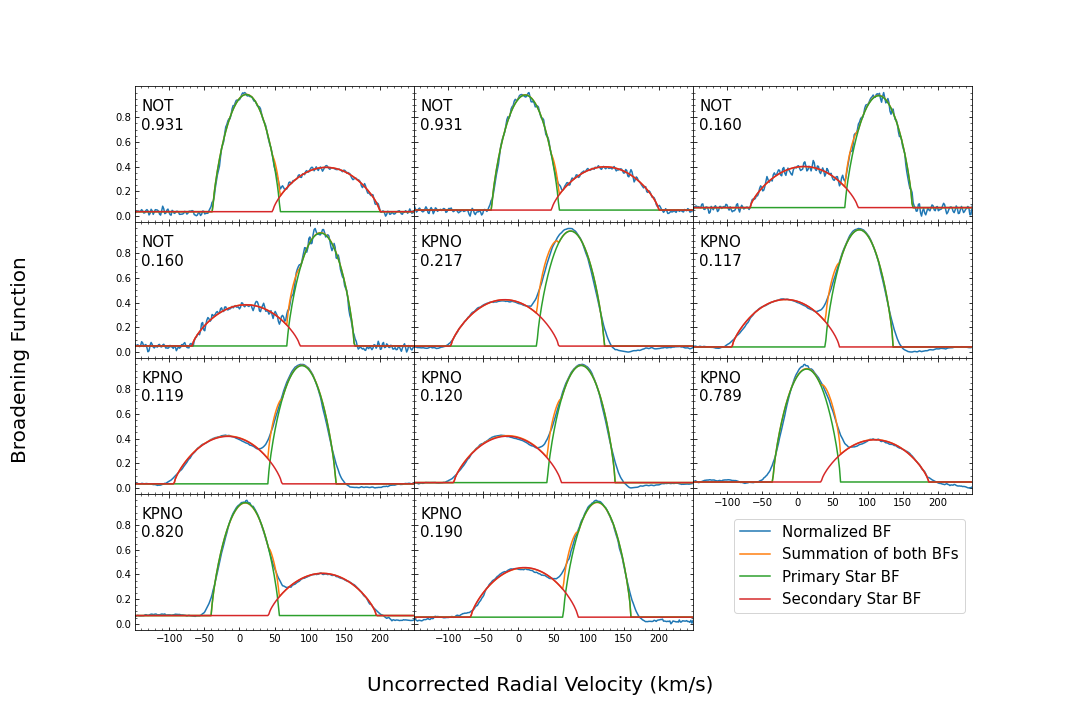}
    \caption{Broadening profiles for the spectra with the most distinguishable contributions from the two stars, with orbital phase given in the upper right corner. The lines shows the fits to the primary star profile (green), the secondary star profile (red), the normalized fit (blue) and the summation (orange). The first four panels are from NOT FIES spectra and the last seven panels are from KPNO spectra.}
    \label{figure:3}
\end{figure}

\subsubsection{Rotational Velocity and Luminosity Ratio\label{lrats}}

Figure \ref{figure:3} shows the broadening functions calculated from the spectra with the cleanest separation of the two stars' peaks. These eleven spectra were taken close to velocity extremes, and gave very consistent measurements for the widths of the rotational broadening function fits (Table \ref{tab:bf}). We derived weighted averages of the primary and secondary star's velocity widths from these spectra: $v_A \sin i = 48.0\pm0.2$ \kms and $v_B \sin i = 76.5\pm0.2$ km s$^{-1}$. The stars are probably rotating considerably slower than single stars of similar brightness in the cluster (see Table \ref{tab:rotvel}). If the stars have aligned rotation axes, then the system cannot be close to synchronism because the primary has a lower measured rotational velocity.

\begin{deluxetable}{cccccc}
\tabletypesize{\footnotesize}
\tablecaption{Rotational Velocity and Luminosity Ratio Measurements for \epc\label{tab:bf}}
\tablehead{\colhead{mJD\tablenotemark{a}} & \colhead{Phase $\phi$} & \colhead{$v_A \sin i$ (km/s)} & \colhead{$v_B \sin i$ (km/s)} & \colhead{$L_B / L_A$} & \colhead{Source}}
\tablecolumns{4}
\startdata
48638.824 & 0.217 & $47.23\pm0.08$ & $76.26\pm0.22$ & 0.680 & KPNO\\
49021.874 & 0.117 & $48.83\pm0.08$ & $75.31\pm0.19$ & 0.642 & KPNO\\
49021.931 & 0.119 & $48.94\pm0.08$ & $75.56\pm0.22$ & 0.639 & KPNO\\
49021.980 & 0.120 & $48.43\pm0.08$ & $75.27\pm0.25$ & 0.628 & KPNO\\
49080.622 & 0.789 & $48.77\pm0.07$ & $77.56\pm0.22$ & 0.593 & KPNO\\
49081.706 & 0.820 & $48.86\pm0.08$ & $76.29\pm0.25$ & 0.595 & KPNO\\
49094.698 & 0.190 & $48.29\pm0.08$ & $76.57\pm0.23$ & 0.683 & KPNO\\
58960.375 & 0.931 & $47.05\pm0.07$ & $77.41\pm0.26$ & 0.604 & NOT FIES\\
58960.379 & 0.931 & $47.05\pm0.07$ & $77.49\pm0.27$ & 0.600 & NOT FIES\\
58968.404 & 0.160 & $47.70\pm0.07$ & $77.02\pm0.21$ & 0.590 & NOT FIES\\
58968.408 & 0.160 & $47.15\pm0.07$ & $76.96\pm0.21$ & 0.585 & NOT FIES\\
\enddata
\tablenotetext{a}{mJD = HJD - 2400000}
\end{deluxetable}

As we will find in section \ref{seds}, the stars of \epc \ have similar surface temperatures, and in that situation, the broadening function areas for the two stars give an indication of the relative luminosities. We therefore used our fits of the spectra in Table \ref{tab:bf} to estimate the luminosity ratio of the two stars in the visible wavelength range, complementing measurements in the infrared from interferometry below. We restricted the rotational velocity fitting to a $\pm 1\sigma$ range around the weighted averages determined above, and redid the fits for the BF heights in order to derive the areas. The results are given in Table \ref{tab:bf}, where the average ratio (with the error in the mean) is $0.622\pm0.010$. This is in agreement with the infrared ratios (see Table \ref{tab:chiparams}), supporting the idea of very similar surface temperatures.

\subsection{Interferometry}\label{interf}

Although lunar occultations of \epc \ have been observed several times \citep{occult1,occult2}, the system was not previously resolved. As we will see from the modeling below, this is largely due to a confluence of aggravating factors: a high inclination, an orientation of the orbit that strongly foreshortens the major axis, and a mostly north-south alignment of the minor axis (putting it fairly close to perpendicular to the ecliptic). These issues have made interferometric observations necessary to resolve the orbit.

Our interferometric observations were obtained from the Center for High Angular Resolution Astronomy (CHARA) at Mt. Wilson Observatory using two different beam combiners called CLIMB and MIRC-X \citep{2005ApJ...628..453T}. The CLIMB beam combiner is a broad-band, single spectral channel instrument that optimizes sensitivity, and is a three-beam expansion of the CLASSIC two-beam combiner \citep{2013JAI.....240004T}. The MIRC-X instrument is an image-plane combiner that simultaneously combines the light from up to six telescopes, and works well on faint binary companions \citep{2020AJ....160..158A}. In the low-resolution ($R = 50$) configuration we used, MIRC-X observes in eight different spectral channels \citep{2018SPIE10701E..23K}. 
The CLIMB data were reduced using the pipeline developed by J.D. Monnier, with the general method described in \citet{monnier} and extended to three-beams (e.g., \citealt{kluska}).  The MIRC-X data were reduced using the standard MIRC-X pipeline version 1.3.5 \footnote{ https://gitlab.chara.gsu.edu/lebouquj/mircx\_pipeline}.

The log of the observations is given in Table \ref{iobserv}, along with the number of fringe visibility (two-telescope interference) and closure phase (derived from the phase shifts in the fringe patterns for three-telescope groups) measurements. Fringe visibilities as a function of projected baseline provide information about the separation and orientation of the two stars in the binary, as well as their relative brightnesses. Visibilities are calibrated using observations on the same night of unresolved single stars. The calibrators are listed in the table, and the observed visibilities were compared to predicted visibilities (based on model angular diameters for a uniformly illuminated disk; \citealt{jmmcdiam}) to derive the normalization for \epc \ observations. The assumed diameters for the listed calibrators can be found in the JMMC Stellar Diameters Catalog V2 \citep{2017yCat.2346....0B}. To ensure that the calibrators are unresolved and fringe visibility is at its maximum for atmospheric conditions, those with diameters less than 0.5 mas were chosen. In almost all cases, calibration observations were taken before and after observations of \epc. In the CLIMB observations, visibilities were measured simultaneously in several combined output beams for the same pairs of telescopes, and we treated the resulting visibility measurements as independent.
Closure phases are observables that provide information about asymmetry in the light distribution, and can provide more powerful measures of the luminosity ratio for the stars in a binary.

Our CLIMB observations were conducted in $K$ band, and our MIRC-X observations in $H$ band (in 8 subchannels), so they give us relative luminosity information for the component stars in the two wavelength regions. Most of our CLIMB observations were affected by marginal weather that made position measurements difficult, but the interferometry still provides observational limits of value to the orbit determination.

\begin{deluxetable}{ccccccl}
\tablecaption{Log of Interferometric Observations of \epc\label{iobserv}}
\tablehead{\colhead{UT Date} & \colhead{mJD Start\tablenotemark{a}} & \colhead{Combiner} & \colhead{$N_{tel}$} & \colhead{$N_{vis}$} & \colhead{$N_{clos}$} & \colhead{Calibrators}}
\startdata
2017 Jan 31 & 57784.77 & CLIMB & 3 & 9 & 3 & HD 73785, 73819\\
2018 Feb 05 & 58154.76 & CLIMB & 3 & 42 & 6 & HD 72779, 74379\\
2018 Feb 06 & 58155.86 & CLIMB & 3 & 41 & 6 & HD 72779, 74379\\ 
2018 Nov 19 & 58441.99 & CLIMB & 3 & 21 & 3 & HD 72779\\
2018 Nov 25 & 58448.02 & CLIMB & 3 & 14 & 2 & HD 72779\\
2018 Dec 13 & 58465.94 & CLIMB & 3 & 28 & 4 & HD 73344, 73533, 74379\\
2018 Dec 24 & 58476.85 & CLIMB & 3 & 26 & 4 & HD 73344\\
2019 Nov 08 & 58795.85 & MIRC-X & 5 & 512 & 448 & HD 72779, 73533, 74379\\
2019 Dec 13 & 58830.98 & MIRC-X & 5 & 272 & 184 & HD 72779\\
2021 Feb 24 & 59269.78 & MIRC-X & 5 & 320 & 320 & HD 72779, 73533\\
2021 Mar 25 & 59298.75 & MIRC-X & 5 & 320 & 320 & HD 72779\\
2021 Mar 29 & 59302.69 & MIRC-X & 6 & 480 & 640 & HD 72779, 74379, 82394\\
2021 Apr 03 & 59307.67 & MIRC-X & 6 & 480 & 640 & HD 72779, 82394, 74379\\
\hline
Total & & & & 2496 & 2570\\
\enddata
\tablenotetext{a}{mJD = HJD - 2400000 at start of observation sequence.}
\end{deluxetable}

\subsection{Spectral Energy Distributions}\label{seds}

\epc \ does not have eclipses, and so to get additional information on
the component stars, we need to utilize photometry. There is an
abundance of calibrated photometry for the binary that we use below to
characterize the spectral energy distribution (SED) of the stars.

{\it Ultraviolet:} The ultraviolet is particularly important for
characterizing A-type stars at the turnoff of the Praesepe cluster.
We obtained photometry from the {\it Galaxy Evolution Explorer} ({\it 
GALEX}; \citealt{galex}) archive in the FUV ($1344-1786$ \AA)
passbands. The cluster was imaged in the FUV for 3329 s as part of a
guest investigator program (GI1 proposal 48,
P.I. M. Burleigh). \citet{galexcal} describes the characteristics of
the GALEX photometry and its calibration to flux. GALEX magnitudes are
on an AB system \citep{oke}, and we used the zero point magnitudes
($m_{FUV} = 18.82$) and reference fluxes ($1.40\times10^{-15}$ erg
s$^{-1}$ cm$^2$ \AA$^{-1}$, respectively) to convert to flux.

The cluster was observed by the Astronomical Netherlands Satellite
(ANS) in five ultraviolet bands centered at 1550, 1800, 2200, 2500,
and 3300 \AA. Magnitudes and measurement errors were taken from
\citet{ans}, and are directly related to monochromatic fluxes. The
absolute flux calibration is described in \citet{anscal}.

{\it Optical, Near Ultraviolet, and Near Infrared:} There are several sources for narrow-
and broad-band photometry in and near the optical range.
We obtained narrow-band Str\"{o}mgren $uvby$ photometry
for cluster stars from \citet{paunzen}, and we employed reference fluxes from
\citet{gray-strom} to convert the magnitudes to fluxes. \citet{vilnius} presented seven-filter Vilnius photometry of cluster
stars, and we converted the reported magnitudes to fluxes using the
zeropoints from \citet{mvb}. Observations in the Geneva system were
taken from the Catalogue of Stars Measured in the Geneva Observatory
Photometric System \citep[4th edition]{geneva}. 
Observations in the
Johnson 13-color system can be found in \citet{13color}, and reference
fluxes for the conversion from magnitudes to fluxes were taken from
the Spanish Virtual Observatory (SVO) Filter Profile Service
\citep{svo1,svo2}.  Additionally, there are spectrophotometic
observations of many bright Praesepe stars in 16 very narrow passbands
in \citet{cb97}. These observations are tabulated as colors relative to
measurements in their 5556 \AA \ filter.

Photoelectric photometry in Johnson $UBVRI$ filters were taken from \citet{ubvri}
and the APASS survey \citep[Data Release 10]{apass10}, which also included Sloan $g^\prime r^\prime i^\prime$ filters.          
The $UBVRI$ measurements were converted to fluxes using reference
magnitudes from Table A2 of \citet{bcp}, accounting for the known
reversal of the zero point correction rows for $f_\lambda$ and
$f_\nu$.
Observations on the four-color $WBVR$ system came from \citet{wbvr}, with flux calibration again making use of \citet{mvb} references.
Photometry in the Tycho filters $B_T$ and $V_T$ was taken from the Tycho Reference Catalogue \citep{tycho}, with
absolute flux calibration information taken from \citet{mvb}.

Due to the brightness of the binary and other cluster members, 
there are only $u$, $g$, and $r$ measurements in the Sloan
Digital Sky Survey (SDSS; \citealt{sdss}).  We used PSF magnitudes
from Data Release 14, calibrated according to \citet{sdsscal}. The
SDSS is nearly on the AB system, with small offsets in $u_{SDSS}$ and
$z_{SDSS}$ that we have also corrected for here. The APASS $g^\prime r^\prime i^\prime$ filters are on an AB system similar to the SDSS, and were calibrated the same way.

Finally, {\it Gaia} has already produced high-precision photometry extending
far down the main sequence of the cluster as part of Early Data Release
3. We obtained the fluxes in the $G$, $G_{\rm BP}$, and $G_{\rm RP}$
bands from the Gaia Archive.

{\it Infrared:} We have obtained Two-Micron All-Sky Survey (2MASS;
\citealt{2mass}) photometry in $JHK_s$ from the All-Sky Point Source
Catalog, and have converted these to fluxes using reference fluxes for
zero magnitude from \citet{2masscal}. We also used photometry in all
four bands ($W1$, $W2$, $W3$, $W4$) from the Wide Field
Infrared Explorer (WISE; \citealt{wise}), which were also converted to
fluxes using tabulated reference fluxes at zero magnitude.
An additional measurement in the S9W filter at 8.22 $\mu$m was taken
from AKARI satellite IRC all-sky survey data \citep{akari}.

Some clues to the CMD positions of the two stars in the binary come from luminosity ratios derived from the broadening function fits in the visible part of the spectrum (see section \ref{lrats}) and from the interferometry in the infrared $H$ and $K$ bands. These ratios all return values near 0.6, implying very similar temperatures and clearly restricting the stars to the upper part of the main sequence. If the spectroscopically-measured luminosity ratio applies to the Gaia $G$ magnitude, then the two stars of \epc \ should have $G = 6.784\pm0.007$ and $G = 7.324\pm0.011$. The primary should therefore be among the stars at the bright end of the CMD gap ($6.8 < G < 7.4$), and the secondary should be approximately 0.2 mag brighter than the faint end of the gap.

To put further limits the components of the binary, we can check whether existing cluster members can act as proxies, such that the binary's SED is reproduced when the SEDs of the two stars are added together.
To find the best proxy for \epc \ A, we considered the five cluster objects that are closest in magnitude to \epc \ A (HD 73210, 73575, 73712, 73785, 73819; see Fig. \ref{sedcomp}). As discussed in Appendix \ref{app_pra}, we first looked for signs of binarity, which would affect the CMD position. The two redder stars (HD 73575 and the possible triple HD 73712) are too red to reproduce the color of \epc \ in combination with a main sequence companion. One of the remaining stars grouped near $(G_{BP}-G_{RP})=0.27$ 
(HD 73210) is a double-lined spectroscopic binary with a fairly extreme mass ratio ($q \approx 0.35$; \citealt{aandw}). The final two stars (HD 73785 and 73819) are the best candidates for a proxy star.

To find the best proxy for \epc \ B, we examined stars at the bright end of the main sequence. We first checked for binarity, and considered whether they could be blue stragglers that are beyond the end of the main sequence in this cluster. The three brightest main sequence stars (HD 72846, 72942, and 73711) are discussed in Appendix \ref{app_pra}. Two of the stars (HD 72942 and 73711) show chemical peculiarities that are often associated with binarity, but only HD 72942 has a clearly detected companion. 

We added the SEDs of different bright and faint proxy star candidates together to see if they could reproduce the SED of the binary, and found that the combination of HD 73819 and 73711 agreed best. Because the spectroscopic luminosity ratio implies that the fainter component of \epc \ B is brighter than any of the cluster turnoff stars, we scaled the SED of HD 73711 upward by 10\% to account for this. Figure \ref{sedcomp} shows the best match. The lower panel shows that we are able to reproduce the binary's SED to within about 5\% in filter bands stretching from the ultraviolet far into the infrared. Redder stars at the brightness level of \epc \ A produce stronger mismatches as a function of wavelength, and can be ruled out. We believe this provides sufficient localization of the two stars to make some use of CMD positions as part of our analysis of \epc. 

\begin{figure}
\plotone{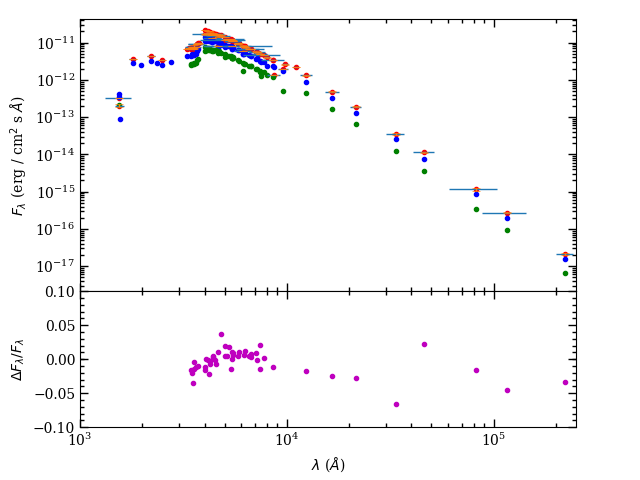}
\caption{{\it Top:} SED for the binary \epc \ from photometry (red
  points), and for the best-fit main sequence stars (HD 73819 with blue, and HD 73711 with green
  points increased in flux by a factor of 1.1).  Horizontal error bars represent the effective width of the
  filter. {\it Bottom:} Fractional difference between the binary star
  photometry fluxes and the best-fit sum of main sequence
  stars.\label{sedcomp}}
\end{figure}

\begin{figure}
\plotone{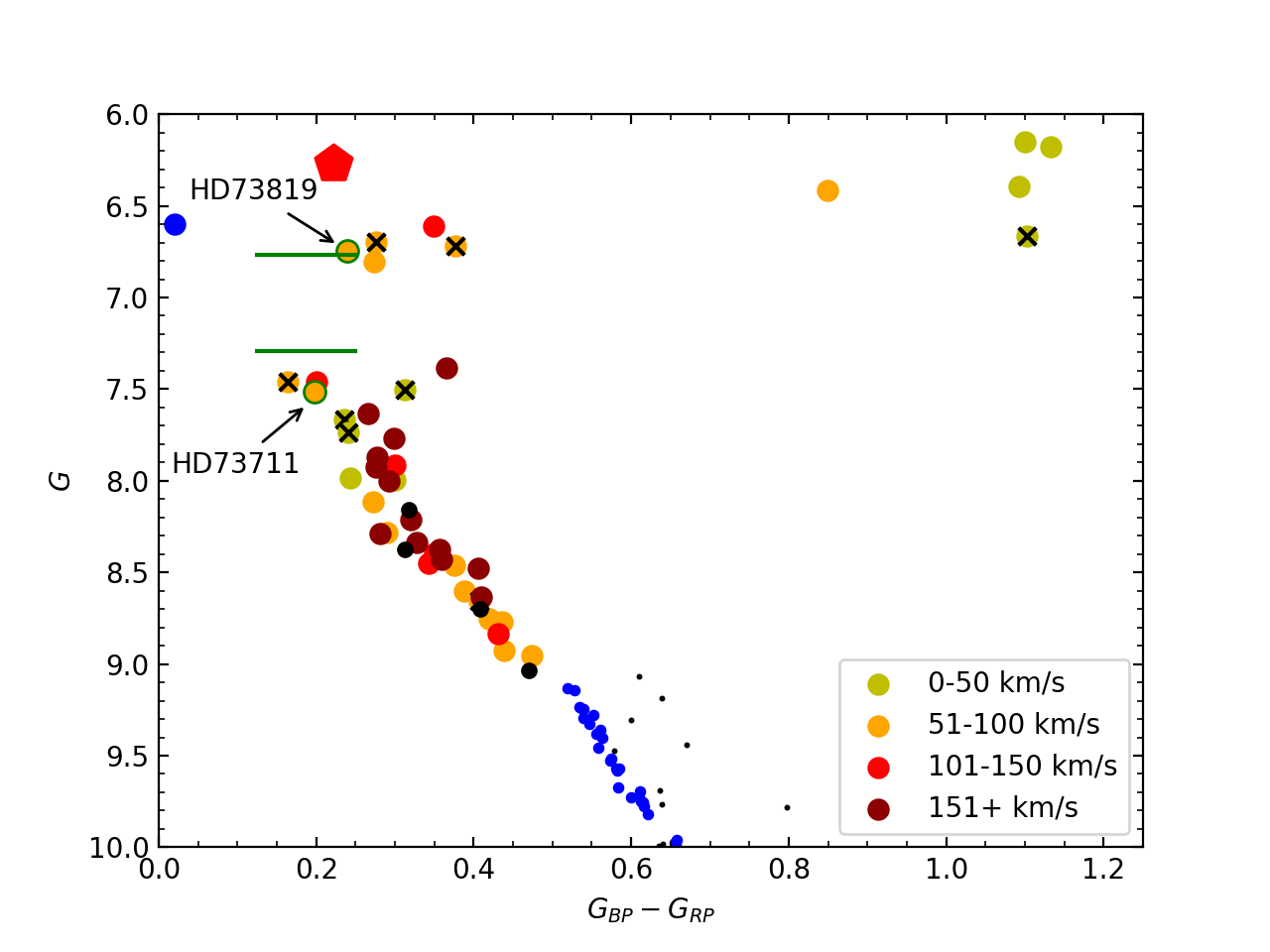}
\caption{Gaia EDR3 color-magnitude diagram for bright Praesepe members. \epc \ is shown with a large red pentagon, the best-fit main sequence stars (HD 73819 and HD 73711) are outlined in green, the blue straggler candidate HD 73666 is a large blue point, likely single faint main sequence stars are small blue points, faint binary candidates are small black points, and stars without measured rotational velocities are large black points. Otherwise the bright stars are color coded by rotational velocity. $\times$ markers indicate binary systems. The green horizontal lines are inferred brightnesses for \epc \ A and B. \label{gcmd}}
\end{figure}

As shown in Figure \ref{gcmd}, the brighter of the proxy stars (EP Cnc or HD 73819) falls among a few stars that appear to be single stars near core hydrogen exhaustion, while the fainter proxy star (HD 73711\footnote{Coincidentally, HD 73711 is cited as a distant companion to \epc \ in the Washington Double Star Catalog \citep{wds}. We cannot ascribe any significance to the apparent similarity between HD 73711 and the fainter component of \epc \ though.}) appears to be close to the populated bright end of the main sequence. Clustering of stars in the CMD may indicate an evolutionary slowdown. Thus there is an elevated probability that the primary star of \epc \ would be in such a phase.

\begin{figure}
\plotone{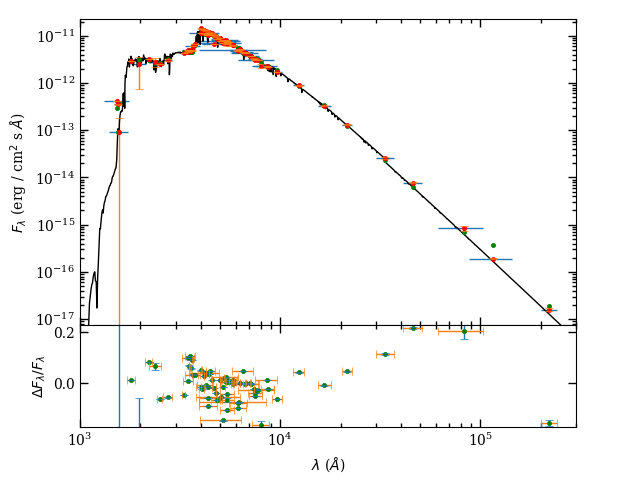}
\caption{{\it Top:} SED for HD 73819 from photometry (red
  points), compared to an ATLAS9 model with $\teff=8060$ K and $\log g = 3.6$ and synthetic photometry calculated from the model (green
  points).  Horizontal error bars represent the effective width of the
  filter. {\it Bottom:} Fractional difference between the observed
  photometry fluxes and best-fit synthetic photometry.\label{hd73819sed}}
\end{figure}

To get an indication of the properties of \epc \ A, we fit the photometric SED of HD 73819 with ATLAS9 models \citep{atlas9}\footnote{Models were calculated using the ATLAS9 fortran code that employed updated 2015 linelists and used temperatures between the published gridpoints.}, as shown in Fig. \ref{hd73819sed}. We used models with fixed values [Fe/H]$=+0.2$ and $\log g = 3.6$, where the surface gravity was chosen from a MIST evolutionary track because there is some sensitivity to gravity in the ultraviolet shortward of the Balmer jump. The best fit temperature (8060 K) was chosen based on the minimization of trends in the fractional residuals (bottom panel) as a function of wavelength. The fit returns a bolometric flux $F_{bol} = 5.39 \times 10^{-8}$ erg cm$^{-2}$ s$^{-1}$ when corrected for wavelength-dependent reddening using the relations from \citet{ccm}. At the distance of Praesepe, this corresponds to a luminosity $L = 59.8\Lsun$.
These values are likely to be fairly good representations of \epc \ A because HD 73819's $G$ magnitude is a good match for the prediction from the spectroscopic luminosity ratio and because the SED of \epc \ is well matched with HD 73819's SED (and it is the biggest contributor to the SED). There are likely to be systematic uncertainties involved, and for this reason, we do not quote uncertainties on these values. 

\begin{figure}
\plotone{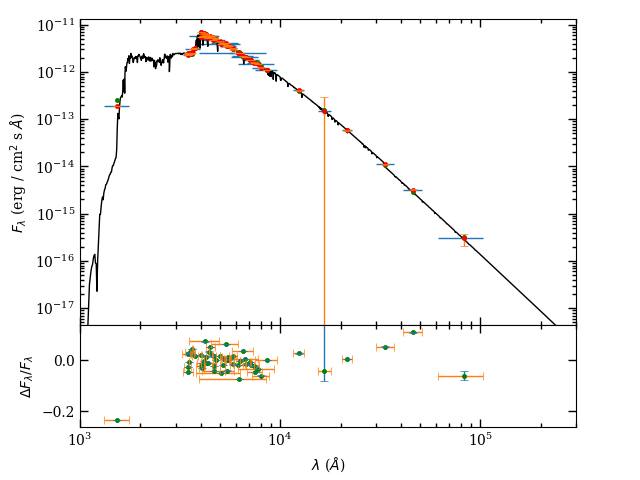}
\caption{{\it Top:} SED for HD 73711 from photometry (red
  points), compared to an ATLAS9 model with $\teff=8330$ K and $\log g = 4.0$ and synthetic photometry calculated from the model (green
  points).  Horizontal error bars represent the effective width of the
  filter. {\it Bottom:} Fractional difference between the observed
  photometry fluxes and best-fit synthetic photometry.\label{hd73711sed}}
\end{figure}

For \epc \ B, we fit the photometric SED of HD 73711 with ATLAS9 models fixing the values [Fe/H]$=+0.2$ and $\log g = 4.0$, as shown in Fig. \ref{hd73711sed}. The star has less ultraviolet data, but the optical part of the spectrum is still well-sampled. The best fit temperature was approximately 8330 K, with bolometric flux $F_{bol} = 2.73 \times 10^{-8}$ erg cm$^{-2}$ s$^{-1}$. At Praesepe's distance, this corresponds to luminosity $L = 30.3\Lsun$. 
In this case, the luminosity is likely to be a significant underestimation of the luminosity of \epc \ B based on the spectroscopic luminosity ratio. Scaling the SED of HD 73711 upward by 10\% in all bands improves the matching of the \epc \ SED when added with that of HD 73819.

Using the bolometric fluxes and effective temperatures, we estimate the angular diameters of the two stars using $\theta^2 / 4 = F_{bol} / \sigma \teff^4$. We find $\theta_1 = 0.194$ mas (from the data for HD 73819) and $\theta_2 = 0.136$ mas (from the data for HD 73711, including the 10\% flux correction).

\section{Binary Star Modeling} \label{sec:process}

With the observational data in hand, the next task is to fit the orbit of \epc \ and determine its parameters. As a first step, we conducted fits to the radial velocity data alone, partly to derive more realistic {\it a posteriori} measurement uncertainties using scatter around the best fits models and partly to look at velocity offsets between datasets. We used the Eclipsing Light Curve code \citep[ELC;][]{elc} to fit eleven parameters: period $P$, reference time of a periastron passage $t_P$, orbit velocity amplitude of the primary star $K_1$, mass ratio $q = M_2 / M_1$, eccentricity $e$, argument of periastron $\omega_1$, and system velocities $\gamma_i$ for five different subsets of data (velocities from KPNO, OHP, and NOT spectra; and literature velocities from \citealt{aandw} and \citealt{abt}). Based on the initial fit using {\it a priori} measurement errors, we scaled the uncertainties on each velocity dataset to return a reduced $\chi^2$ value of approximately 1, and then we refitted. These results are reported in Table \ref{tab:chiparams}. The system velocities were in reasonable agreement, but there appeared to be significant offsets that we then corrected for before the joint interferometry-spectroscopy fitting. (The NOT zeropoint was used as the reference.) We find $\Delta \gamma(\mbox{KPNO}-\mbox{NOT})=-1.32\pm0.08$ km s$^{-1}$, $\Delta \gamma(\mbox{OHP}-\mbox{NOT})=-0.48\pm0.15$ km s$^{-1}$, $\Delta \gamma(\mbox{MWO}-\mbox{NOT})=-0.34\pm1.09$ km s$^{-1}$, and $\Delta \gamma(\mbox{A\&W99}-\mbox{NOT})=-2.06\pm0.15$ km s$^{-1}$.

For the fits to the combined radial velocity and interferometry (visibility amplitudes and closure phases) dataset, we developed a program (which we call {\it interfRVorbit}). In this case, we use a set of thirteen parameters to model the orbits: period $P$, reference time of a periastron passage $t_P$, orbit velocity amplitudes of both stars $K_1$ and $K_2$, systemic velocities for both stars $\gamma_1$ and $\gamma_2$, eccentricity $e$, argument of periastron $\omega$, argument of the ascending node $\Omega$, inclination $i$, angular size of the binary semi-major axis $a$, and brightness ratios in $H$ and $K_s$ bands. The interferometric observations do not have much leverage on the unresolved angular diameters of the stars $\theta_1$ and $\theta_2$, and so we fix these at the values calculated from the SEDs (in the previous section). Observable quantities are then calculated from these parameters through a forward model. In the case of the interferometry, the complex visibility for a uniform disk of angular diameter $\theta$ is
\[ V = \frac{2 J_1(\pi B \theta / \lambda)}{\pi B \theta / \lambda} \]
where $J_1$ is the first-order Bessel function and $B$ is the projected baseline of the interferometer at the star's sky position.
For a binary system, the observable squared visibility is calculated from
\[ V^2 = \frac{V_1^2 + r^2 V_2^2 + 2 \left| V_1 \right| \left| V_2 \right| r \cos \left[ 2\pi \vec{B} \cdot \vec{s} / \lambda\right]}{(1 + r)^2} \]
where $V_1$ and $V_2$ are the complex visibilities of the individual stars, $r$ is the luminosity ratio in the observed band, $\vec{s}$ is the vector of the angular separation of the two stars on the sky, and $\vec{B}$ is the baseline vector for the interferometer. For the fits, we have assumed uniformly illuminated disks for both stars, and used a single luminosity ratio for all observations within a bandpass (in our case, $H$ or $K_s$). Bandpasses are broken into smaller parts ($n = \mbox{int} (200 \Delta \lambda / \lambda)$, with $n \ge 2$) and the calculated visibilities are averaged
\[ V_{band}^2 = \frac{1}{n} \sum_{i=1}^{n} V_i^2 \]
The closure phase can be calculated as the product of the complex visibilities $V$ measured by each pair of telescopes around a closed triangle of baselines. All of the interferometric astrometry measurements are referenced to the brighter star in the binary, so proper motion does not play any role in the fits.

We search the parameter space using a genetic algorithm \citep{charb}.   The goodness of the fit is based on the total $\chi^2$ value from comparing model predictions to observed data. After an initial optimization to find a best fit using {\it a priori} measurement uncertainties, we scaled the uncertainties for each observational dataset (interferometric visibilities, closure phases, and radial velocities separated by measured star and spectroscopic source) to return a reduced $\chi^2$ of approximately 1. In this way, we make the datapoint uncertainties consistent with the observed scatter around the best-fit model and attempt to produce appropriate weightings for the different data types in subsequent fits.

We subsequently conducted a 20000 generation run (with approximately 100 model members per generation) to seek the best-fit model and to derive the parameter uncertainties from trial models near the minimum. The best-fit parameters are given in Table \ref{tab:chiparams}, where the spectroscopic orbit from \citet{aandw} is also provided for comparison. Uncertainty estimates were generated from the full parameter range covered by models with $\chi^2$ within 1 of the minimum \citep{avni}.

\begin{deluxetable}{c|cc|cc|cc}
\tablecaption{Orbit Fitting Results for \epc}
\tablehead{& \multicolumn{2}{c}{Interf. \& RVs  (Equal L ratios)}  & \multicolumn{2}{c}{RVs Only} & \multicolumn{2}{c}{A\&W99}\\
\colhead{Parameter} & &  \colhead{$ \sigma $} &  & \colhead{$\sigma$} &  & \colhead{$\sigma$}}
\startdata
$P$ (d) & 35.14101 & 0.00005  & 35.14113 & 0.00012 & 35.202 & 0.033\\
$t_P - ~2400000$ & 48314.598 & 0.016 & 48314.56 & 0.04 & 48313.5 & 0.7\\
$\gamma_1$ (km s$^{-1}$)& 34.71 & 0.03 & 34.78\tablenotemark{a} & 0.07 & 29.9 & 1.1\\
$K_1$ (km s$^{-1}$)& 56.60 & 0.03 &  56.53 & 0.08 & 53.0 & 1.9\\
$K_2$ (km s$^{-1}$)& 61.55 & 0.10 & 61.50 & 0.08 & 67.8 & 3.9\\
$q = M_2/M_1$& 0.9196 & 0.0015 & 0.9192 & 0.0005 & 0.78\tablenotemark{b} & \\
$e$ & 0.4195 & 0.0003 & 0.4165 & 0.0021 & 0.32 & 0.04\\
$\omega_1 (^{\circ})$ & 258.38 & 0.02 &  258.0 & 0.2 & 265 & 5\\
$\gamma_2$ (km s$^{-1}$) & 34.88 & 0.09 & & & &\\
$\Omega (^{\circ})$& 356.069 & 0.014 & & & & \\
$i (^{\circ})$& 81.454 & 0.010 & & & & \\
$L_2/L_1 (H)$ & 0.61200 & 0.0007 & & & & \\
$a$ (mas) & 1.9127 & 0.0004 (0.0047\tablenotemark{c}) & & & & \\
$M_1/M_{\odot}$ & 2.420 & 0.008 & & & & \\
$M_2/M_{\odot}$ & 2.226 & 0.004 & & & & \\
\enddata
\tablenotemark{a}{Value derived from measured \citet{aandw} spectra.}
\tablenotetext{b}{The published value in \citet{aandw} does not follow from their measured $K_1$ and $K_2$, and we recalculate it here.}
\tablenotetext{c}{Systematic error due to absolute wavelength calibration uncertainty for MIRC-X instrument \citep{vega_mirc}.}
\label{tab:chiparams}
\end{deluxetable}

\begin{figure}
    \centering
    \includegraphics[scale=.9]{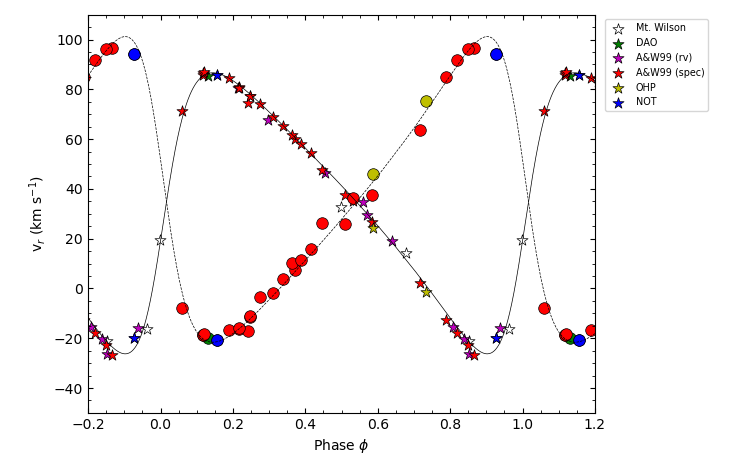}
    \caption{Radial velocity versus orbital phase for the components of \epc. Star symbols represent the primary star and circles represent the secondary star.}
    \label{figure:4}
\end{figure}

Figure \ref{figure:4} compares the radial velocity measurements with the best-fit model as a function of orbital phase. The eccentricity of the orbits are clearly seen. Although we allowed the systematic velocities for the two stars to be different to account for differences in gravitational redshifts or convective blueshifts between the two stars, the velocities disagree by only about 0.2 \kms.

\begin{figure}
    \centering
    \includegraphics[scale=.9]{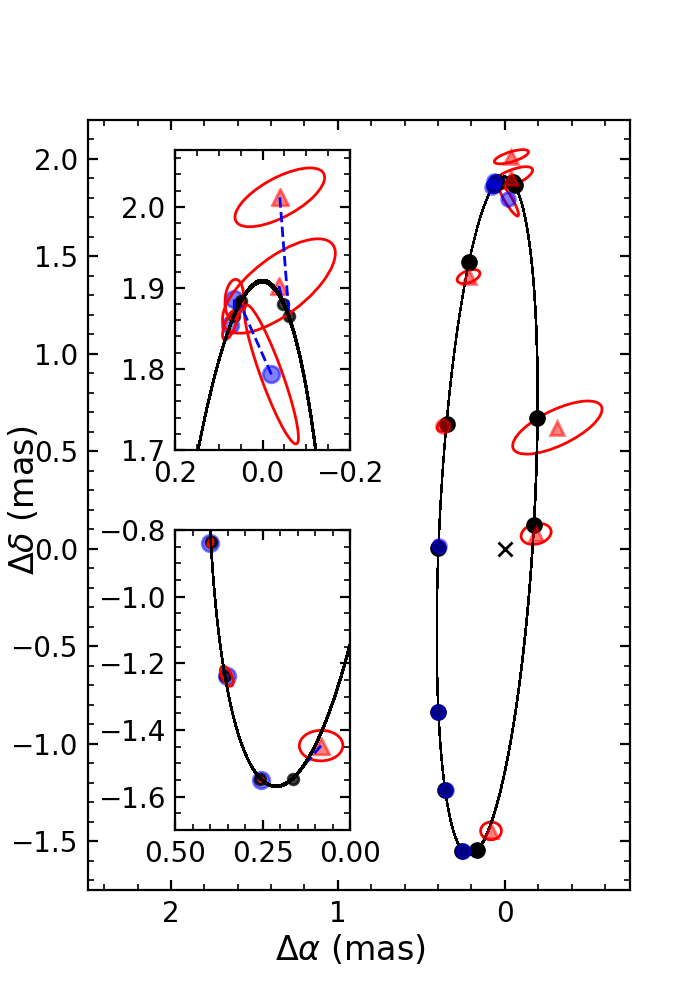}
    \caption{{ Sky orbit for \epc \ B relative to \epc \ A (marked with an $\times$). Black points mark the model orbital position during the measurements with the CLIMB beam combiner at CHARA. Blue points are position measurements from the MIRC-X beam combiner. Red points are position measurements from the CLIMB beam combiner (triangles are values closest to the predicted position, and not the global best binary fit). Red ellipses show 
    $5\sigma$ uncertainties. Secondary star motion is counterclockwise in this diagram. The inset panels are zooms on observations taken near the extremes of the orbit. In the insets, blue dashed lines connect the measured position to the model's predicted position. }}
    \label{figure:5}
\end{figure}

\begin{figure}
    \includegraphics[scale=1.0]{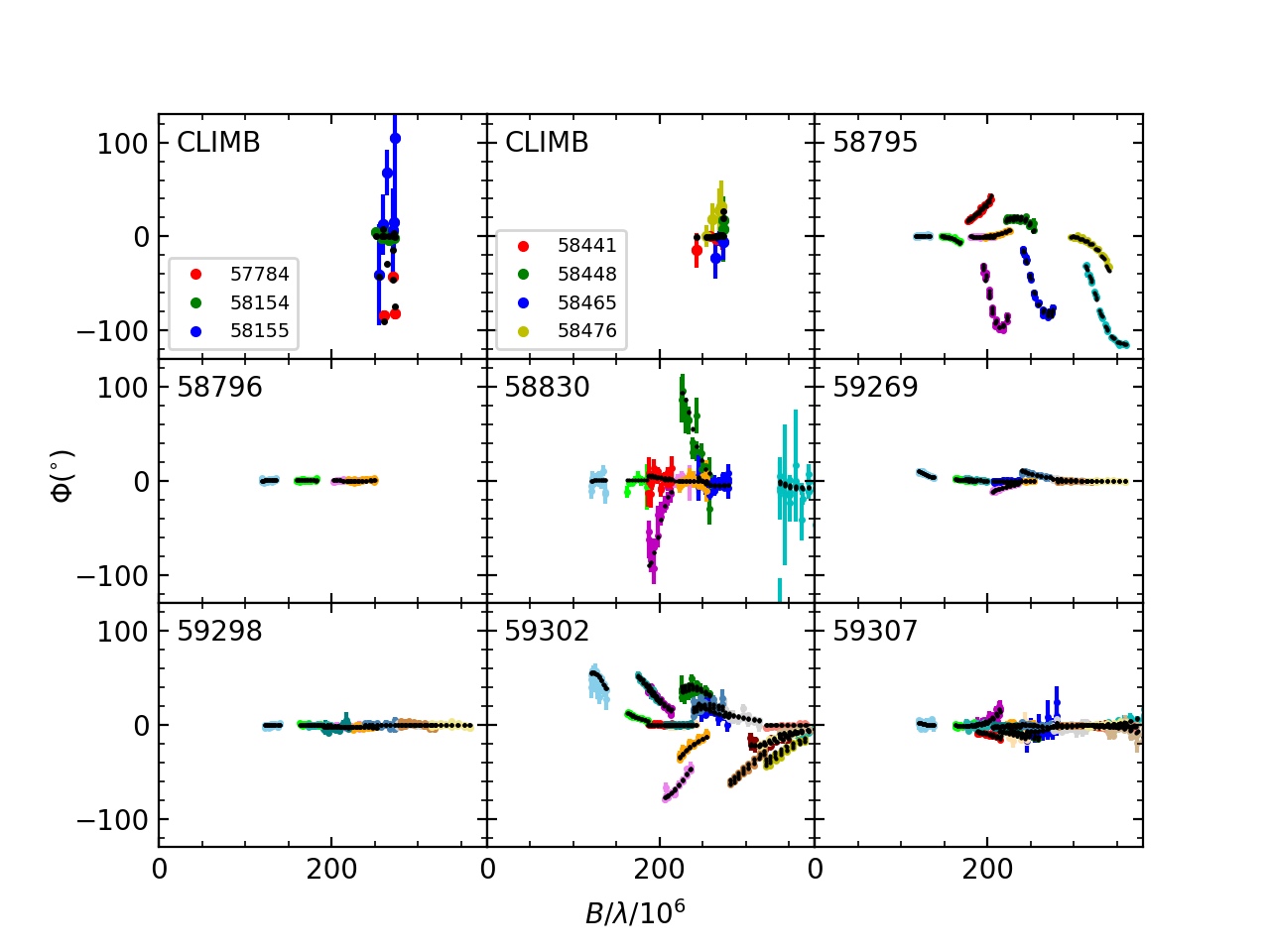}
    \caption{Closure phases ($\Phi$) as a function of the interferometric baseline for CLIMB and MIRC-X data, separated by date of observation. Here the baseline is the sum of the baseline lengths for the first two legs of the triangle of telescopes. Black dots are the predicted values, and colored dots are observed values on different dates (CLIMB) or from different telescope pairs (MIRC-X).}
    \label{figure:7}
\end{figure}

The best fitting model of the binary system's orbit on the sky is shown in Figure \ref{figure:5}. Although we specifically fit interferometric observables (fringe visibilities and closure phases), in the figure we show relative position measurements for the epochs when they could be derived. These position measurements are given in Table 4 and were computed using a binary grid search procedure\footnote{https://www.chara.gsu.edu/analysis-software/binary-grid-search; \citealt{schaefer}}. The position uncertainties include 0.5\% uncertainty in the MIRC-X wavelength scale \citep{2020AJ....160..158A}. These position measurements are given in Table \ref{intpos}.
The comparisons between interferometric observables and model predictions are shown in Figures \ref{figure:6} and \ref{figure:7}. For perspective, the closure phases will be near zero (or $180\degr$) for point-symmetric sources, but the maximum excursions away from zero are good indicators of the luminosity ratio of the stars in the binary. (Larger deviations reflect luminosity ratios closer to 1.)

\begin{figure}
    \includegraphics[scale=1.0]{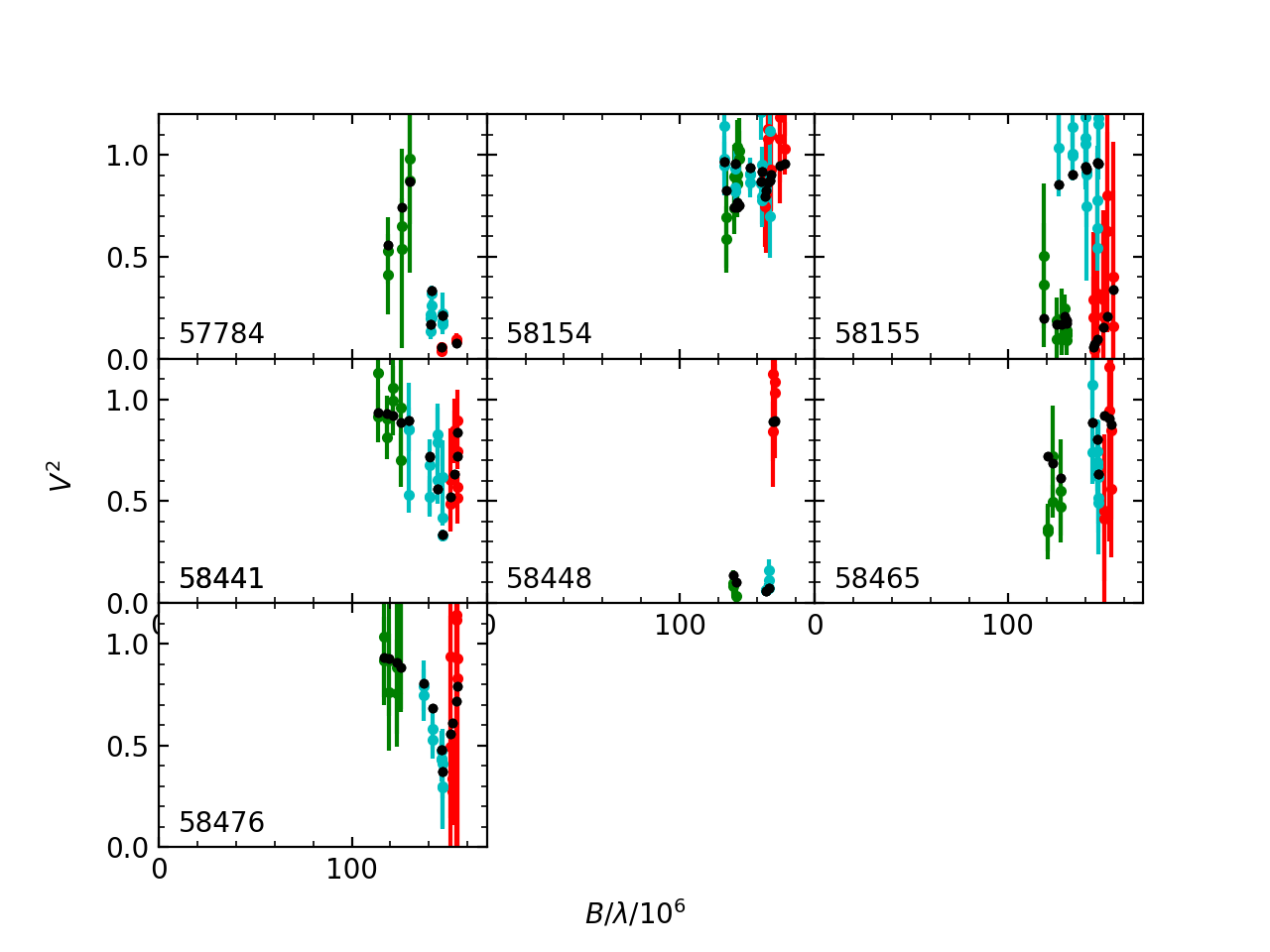}
    \caption{Squared visibilities as a function of interferometric baseline for the CLIMB data, separated by date of observation. Black dots are the predicted values and the colored dots are the observed values. Different colors in the panels represent measurements from different telescope pairs.}
    \label{figure:6}
\end{figure}

\begin{figure}
    \includegraphics[scale=1.0]{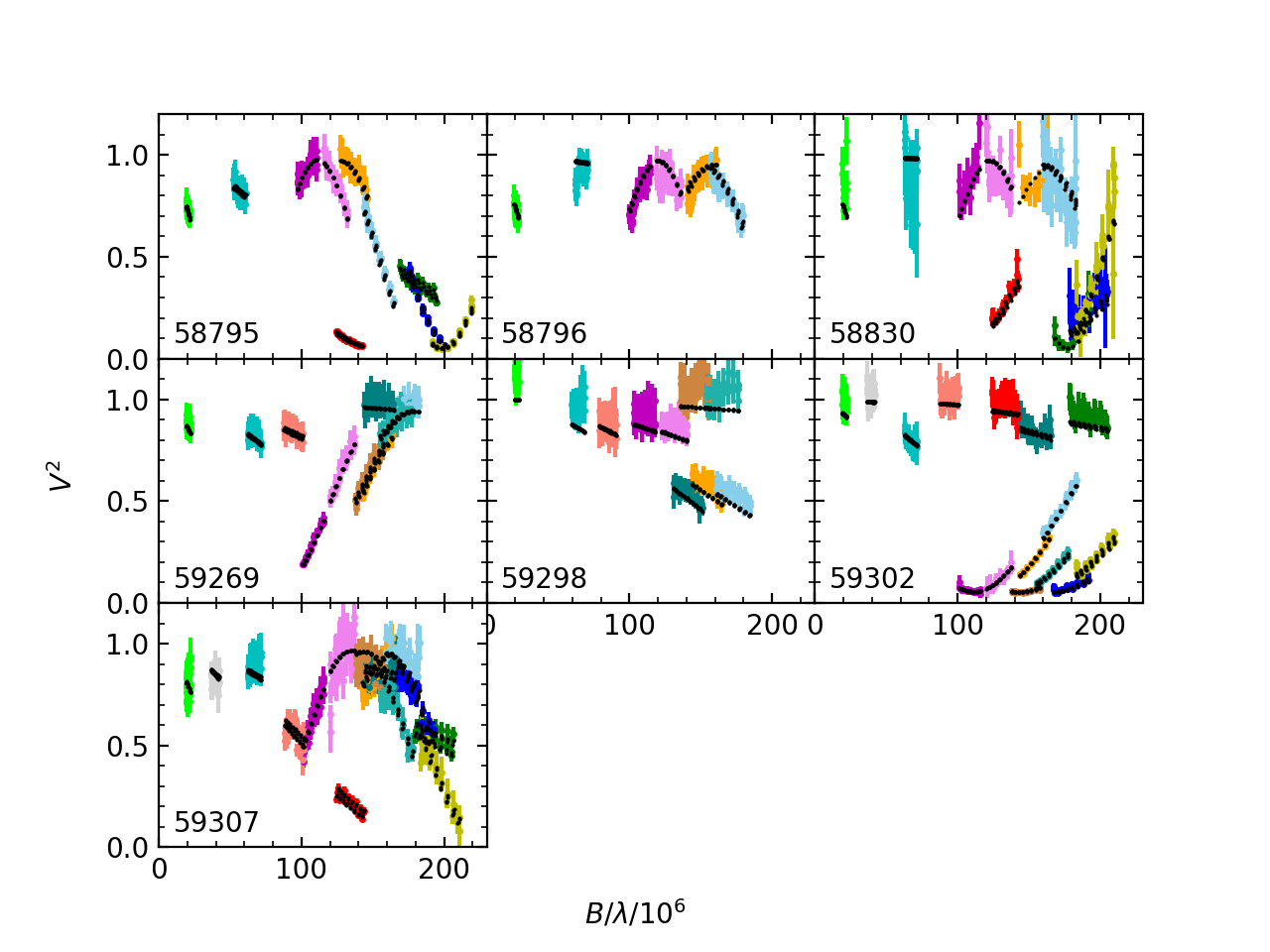}
    \caption{Squared visibilities as a function of interferometic baseline for the MIRC-X data, separated by date of observation. Black dots are the predicted values and the colored dots are the observed values. Different colors in the panels represent measurements from different telescope pairs.}
    \label{figure:6}
\end{figure}

\begin{deluxetable}{crrrrrcr}
\tablecaption{Interferometric Relative Position Measurements for \epc}
\tablehead{\colhead{mJD\tablenotemark{a}} & \colhead{$\rho$} & \colhead{$\theta$} &  \colhead{$\sigma_{maj}$\tablenotemark{b}} & \colhead{$\sigma_{min}$\tablenotemark{b}} & \colhead{$\varphi$\tablenotemark{b}} & \colhead{$f$} & \colhead{$\sigma_f$} \\
& \colhead{(mas)} & \colhead{(deg)} & \colhead{(mas)} & \colhead{(mas)} & \colhead{(deg)} & \colhead{$L_1/(L_1+L_2)$} & }
\startdata
\multicolumn{8}{c}{CLIMB Combiner\tablenotemark{c}} \\
57784.826 & 0.7313 & 30.61 & 0.0138 & 0.0091 & 151.5 & 0.619 & 0.008\\ 
58154.878 & 0.2027 & 292.23 & 0.0364 & 0.0208 & 169.4 & 0.620\tablenotemark{d} & \\ 
58155.860 & 0.6974 & 333.17 & 0.1142 & 0.0372 & 158.4 & 0.619 & 0.122\\ 
58441.988 & 1.9026 & 358.87 & 0.0543 & 0.0159 & 160.9 & 0.620\tablenotemark{d} & \\
58448.033 & 1.4136 & 8.84 & 0.0283 & 0.0125 & 163.5 & 0.620\tablenotemark{d} & \\
58465.939 & 1.4492 & 176.73 & 0.0249 & 0.0183 & 179.3 & 0.620\tablenotemark{d} & \\
58476.882 & 2.0120 & 358.87 & 0.0422 & 0.0104 & 165.7 & 0.620\tablenotemark{d} & \\
\multicolumn{8}{c}{MIRC-X Combiner}\\
58795.986 & 1.8559 &   2.28 & 0.0093 & 0.0041 & 132.3 & 0.613 & 0.001\\ 
58796.057 & 1.7933 & 359.35 & 0.0414 & 0.0087 &  55.1 & 0.572 & 0.052\\ 
58830.984 & 1.8869 &   1.96 & 0.0101 & 0.0082 & 112.3 & 0.617 & 0.008\\ 
59269.783 & 1.2874 & 164.14 & 0.0129 & 0.0057 &  66.2 & 0.617 & 0.012\\ 
59298.759 & 0.3933 &  87.66 & 0.0129 & 0.0085 &  46.6 & 0.641 & 0.021\\ 
59302.695 & 0.9283 & 154.57 & 0.0048 & 0.0023 & 113.6 & 0.617 & 0.002\\ 
59307.677 & 1.5710 & 170.74 & 0.0022 & 0.0011 & 138.5 & 0.608 & 0.011\\ 
\enddata
\tablenotetext{a}{mJD = HJD $-2400000$}
\tablenotetext{b}{The size of the error ellipse major and minor axes, and the orientation $\varphi$ of the major axis relative to north.}
\tablenotetext{c}{CLIMB entries give the binary fit positions most consistent with the orbit.}
\tablenotetext{d}{Luminosity ratio was assumed on nights when the interferometry did not give a reliable independent measurement.}
\label{intpos}
\end{deluxetable}

When we allow the $H$ and $K$ band luminosity ratios to be fit independently, the ratios are in rough agreement, but differ significantly. Because both filters sample the long-wavelength side of the SEDs of these hot stars, it is to be expected that the ratios would be nearly identical. We therefore conducted a second fitting run where we forced the $H$ and $K$ luminosity ratios to be the same. Predictably, the final ratio came out very close to the previous $H$ ratio due to the much greater number of interferometric data in that band. It does also lead to a slight shift in the fitted orbit, however, because the choice affects the observational limits from the CLIMB data.

The masses of the primary and secondary stars were computed from the orbital parameters for each run, and we compute $1\sigma$ uncertainties on the masses using the range of masses producing $\chi^2$ values within 1 of the minimum value. 

Using the period $P$, orbit velocity amplitudes $K_1$ and $K_2$, and inclination $i$ from Table \ref{tab:chiparams}, we can calculate the true semi-major axis of the orbit. Used along with the angular size, we get a distance of $183.1\pm0.2\pm0.5$ pc.

(The second quoted error is systematic, deriving from uncertainty in absolute wavelength calibration of the MIRC-X instrument.) This is in good agreement with the geometric distance calculated using the {\it Gaia} EDR3 parallax for \epc \ ($ d = 184.4^{+1.8}_{-1.4} $ pc; \citealt{bailer}).

\section{Discussion\label{sec:disc}}

We can now use the estimated photometric characteristics of the stars of \epc \ along with the interferometrically measured masses and radii to compare with models. One goal of these comparisons is to derive an age as accurate as possible for the cluster, but we must be wary of the possibility of physics errors in the modeling codes that introduce systematic errors into derived ages. As a result, it is important to examine the cluster stars more broadly in order to validate the coded physics. For our purposes, the treatment of convective cores is critical, as the extent of the core in a star directly determines the amount of fuel available for the main sequence phase. The masses for both stars in \epc \ put them in the plateau of the amount of convective core overshooting embedded into codes, where a fixed overshooting distance (often in terms of pressure scale height $H_P$) is implemented. At lower masses (usually less than $2\msun$), a "ramp" is enforced in which the amount of core overshooting is decreased linearly with decreasing mass down to zero for stars that no longer have convective cores. So in principle, the stars of \epc \ avoid uncertainties related to the implementation of the ramp, but they remain a tough test of convective overshooting because they are stars with precise masses at or near the end of their main sequence lives, where the convective core history produces its most noticeable effects.

In Table \ref{tab:rotvel}, we summarize photometry, rotational velocity, and multiplicity information for the brightest stars in Praesepe. We primarily used {\it Gaia} EDR3 information on parallaxes, positions, and proper motions to construct a likely member list. We selected stars based on a position within $5\degr$ of $(\alpha, \delta) = (130.10, 19.667)$, a proper motion vector within 6 mas yr$^{-1}$ of $(\mu_\alpha, \mu_\delta) = (-36.047, -12.917)$, a parallax within 0.4 mas of 5.371 mas \citep{gaiahrs}. Because there is evidence of tidal tails stretching much farther from the cluster center, we examined the membership list of \citet{rs} for additional bright members. While \citeauthor{rs} do identify some bright high-probability members in the tidal tails, we find that nearly all of them have radial velocities inconsistent with cluster membership (deviating from the cluster mean by 10 \kms or more). All of the stars in the table have systematic radial velocities within 5 \kms of the cluster mean (sources: \citealt{aandw,rm98,debruijne,yang,cummings}).

\begin{deluxetable}{rccccccl}
\tabletypesize{\scriptsize}
\tablecaption{Rotational Velocities and Multiplicity of Bright Stars ($G < 9.0$) in Praesepe}
\tablehead{ \colhead{KW} & \colhead{HD/BD} &  \colhead{$G$} & \colhead{$ v \sin i$ (km/s)}& \colhead{Ref.} & \colhead{Multiplicity} & \colhead{Ref.} & \colhead{Notes}}
\startdata
253 & 73665 & 6.15 & $<10$ & 3 & No & 3,14 & RC\\
283 & 73710 & 6.18 & $<10$ & 3 & No & 3,14,15 & RC\\
212 & 73598 & 6.39 & $<10$ & 3 & speck,occ & 14,16 & RGB? \\
    & 72779 & 6.42 & 99 & 3 & No & 3,14 & SGB\\
428 & 73974 & 6.66 & $<10$ & 3 & SB1O,occ & 3,15 & SGB\\
265 & 73666 & 6.58 & 9/10 & 3,5 & No/occ & 3/14 & BSS\\
204 & 73575 & 6.61 & 150/129/127/140 & 1,3,5,11 & No & 3,14,15,16 & $\delta$ Sct\\
 50 & 73210 & 6.70 & 80/90/90,80 & 1,2,3 & SB2O & 3 & \\
284 & 73712 & 6.72 & 56:,61: & 3 & SB2O & 3,14,16 & $\delta$ Sct, triple?\\
348 & 73819 & 6.74 & 152/180/79 & 1,3,11 & No & 3,14,15,16 & $\delta$ Sct\\
328 & 73785 & 6.80 & 85/91 & 1,3 & No/SB2? & 3,15,16/7 & \\
150 & 73449 & 7.38 & 235/229 & 1,3 & No/SB? & 3,14,15/7 & $\delta$ Sct\\
534 & 72942 & 7.46 & 62/73 & 1,5 & SB1O & 3 & A/Am hybrid\\
    & 72846 & 7.46 & $119\pm6$ & 6 & No? & 14 & \\
229 & 73619 & 7.50 & 20,18/$11.2\pm0.5$ & 3,4 & SB2O & 3,7 & Am\\
276 & 73711 & 7.52 & 60/66/$66\pm11$/62 & 1,3,4,5 & No/SB2? & 3,14,15/7,9 & Am \\
207 & 73576 & 7.63 & 200/204/$208\pm11$ & 1,3,10 & No/SB1? & 3,14/7 & $\delta$ Sct\\
279 & 73709 & 7.67 & 20/$17.3\pm0.3$/10 & 3,4,5 & SB1O & 3,7 & quadruple,Am\\
 40 & 73174 & 7.74 & 11/$7.3\pm1.1$/$<$5 & 3,4,5 & SB1O,SB1O & 3 & triple,Am \\
323 & 73763 & 7.77 & $200\pm3$ & 10 & No? & 14 & $\delta$ Sct\\
449 & 74050 & 7.86 & 150/$188\pm8$/145 & 1,6,11 & No? & 14 & $\delta$ Sct\\
445 & 74028 & 7.92 & 160/$150\pm9$ & 1,6 & No? & 14,15,16 & $\delta$ Sct\\
203 & 73574 & 7.92 & 108/108/$102\pm4$ & 1,3,6 & No/bin & 3/14,15 & \\
286 & 73730 & 7.98 & 30/$34.4\pm3.4$/29/35 & 1,4,5,8 & No? & 14,15 & Am\\
    & 74656 & 7.99 & $25\pm1$ & 6 & No? & 13 & Am\\
385 & 73890 & 8.00 & 165/141 & 1,11 & No? & 14 & $\delta$ Sct\\
114 & 73345 & 8.11 & 98/85/$85\pm4$/90 & 1,2,6,8 & No? & 14,15 & $\delta$ Sct\\
    & 74740 & 8.16 & & & No? & 12,13 & \\
 45 & 73175 & 8.21 & 180/$163\pm9$/175 & 1,6,8 & No? & 14 & $\delta$ Sct\\
143 & 73430 & 8.28 & 82/73/80 & 1,5,8 & No?/occ & 14/15 & \\
375 & 73872 & 8.29 & 180/135 & 1,8 & No?/SB1? & 14,15/7 & $\delta$ Sct\\
    & 74718 & 8.33 & $155\pm7$ & 6 & No? & 17 & \\
    & 72757 & 8.37 & $179\pm10$ & 6 &  & & \\
    & 74720 & 8.38 & & & speck & 18 & no rv measure?\\
    & 74589 & 8.40 & $127\pm5$ & 6 & No? & 13 & pulsator\\
340 & 73798 & 8.43 & 175/$200\pm11$/200 & 1,6,8 & No? & 14 & $\delta$ Sct\\
154 & 73450 & 8.45 & 138/140/$138\pm7$/166 & 1,2,6,8 & No? & 14,15 & $\delta$ Sct\\
    & 74587 & 8.46 & $90\pm4$ & 6 & No? & 13 & $\delta$ Sct\\
429 & 73993 & 8.48 & 200/$240\pm14$/210 & 1,6,8 & No? & 14 & \\
318 & 73746 & 8.60 & 110/87/$95\pm4$/95 & 1,2,6,8 & No? & 14 & $\delta$ Sct\\
 38 & 73161 & 8.63 & 160/190 & 1,8 & No? & 14,15 & \\
350 & 73818 & 8.65 & 85/81/66/65 & 1,4,5,8 & SB1 & 5 & Am\\
    & 70297 & 8.70 & & & & & no rv measure?\\
271 & +20 2161 & 8.76 & 85/69/78 & 1,2,8 & No? & 14 & \\
    & 74135 & 8.77 & $100\pm4$ & 6 &  & & pulsator\\
226 & 73616 & 8.83 & 125/110/115 & 1,2,8 & No? & 14,15 & pulsator\\
124 & 73397 & 8.92 & 100/86/91 & 1,2,8 & SB1O & 7 & pulsator\\
370 & 73854 & 8.95 & 137/68/70 & 1,2,8 & No? & 14,15 & pulsator\\
    & 73620 & 9.04 & & & & & \\
\enddata
\label{tab:rotvel}
\tablenotetext{}{NOTE: KW identifications are from \citet{kw}. References: 1: \citet{mcgee}. 2: \citet{rachford}. 3: \citet{aandw}. 4: \citet{debernardi}. 5: \citet{2007AA...476..911F}. 6: \citet{fossati08}. 7: \citet{rm98}. 8. \citet{cummings}. 9: \citet{bc98}. 10: \citet{royer}. 11: \citet{bushhintz}. 12. \citet{evans85}. 13. \citet{grenier}. 14. \citet{speckle}. 15. \citet{occult1}. 16. \citet{occult2}. 17. \citet{eande83}. 18. \citet{horch}.}
\end{deluxetable}

\begin{figure}
\plotone{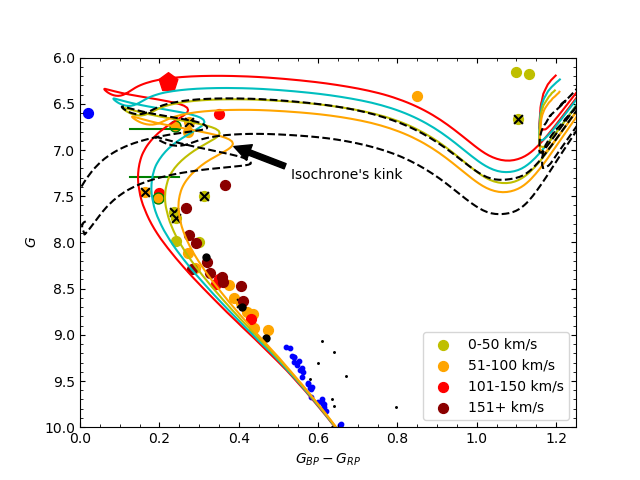}
\caption{MIST isochrones \citep{2016ApJ...823..102C} for ages 600 (red), 650 (cyan), 700 (yellow), and 750 (orange) Myrs. MIST evolutionary tracks are also shown (black) for masses of the primary and secondary star. ($M_p = 2.420 \msun$,~ $M_s = 2.226 \msun$). $\times$ markers indicate binary systems. To graph the evolution tracks and isochrones, a distance modulus $(m-M)_0= 6.33$, initial rotation $v/v_{crit} = 0$, iron abundance [Fe/H]$ = +0.16$, and reddening $E(G_{BP}-G_{RP})= 0.073$ were used.  \label{mist}}
\end{figure}

\begin{figure}
\plotone{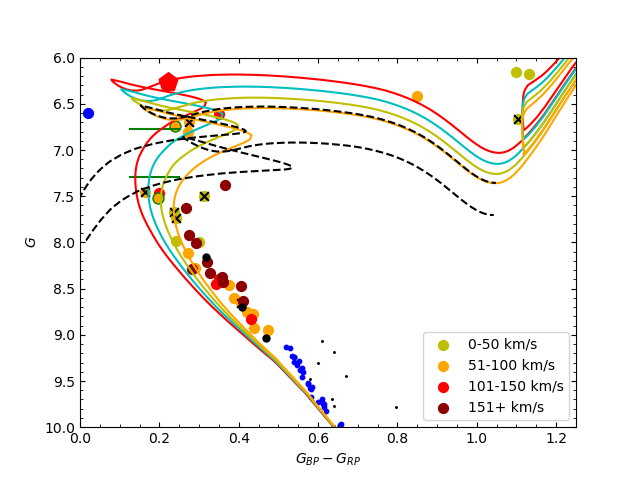}
\caption{BaSTI-IAC isochrones \citep{2018ApJ...856..125H} for ages 600 (red), 650 (cyan), 700 (yellow), and 750 (orange) Myrs. BaSTI-IAC evolutionary tracks are also shown (black) for masses of the primary and secondary stars ($M_p = 2.420 \msun$, $M_s = 2.226 \msun$). $\times$ markers indicate binary systems. In the graph, distance modulus $(m-M)_0 = 6.33$, iron abundance [Fe/H]$ = +0.16$, and reddening $E(G_{BP}-G_{RP})= 0.073$ were used. Overshooting and diffusion were also allowed under available grids. \label{basti}}
\end{figure}

\begin{figure}
\plotone{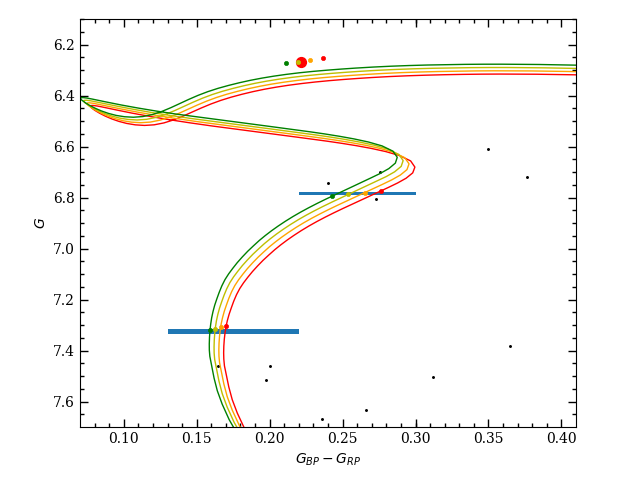}
\caption{MIST isochrones \citep{2016ApJ...823..102C} for ages 630 (green), 635 (light green), 640 (yellow), and 645 (orange) Myr. Distance modulus $(m-M)_0= 6.33$, iron abundance [Fe/H]$ = +0.14$, and reddening $E(G_{BP}-G_{RP})= 0.073$ were used. Colored dots show the model positions of stars with the measured masses of the components of \epc, and blue rectangles show the $G$ magnitude and uncertainty of the components as inferred from spectroscopy.\label{cmd_decomp}}
\end{figure}

Figures \ref{mist} -- \ref{parsec} show comparisons between theoretical isochrones and {\it Gaia} photometry for the cluster. In Figures \ref{mist} and \ref{basti}, evolution tracks for the measured masses of the \epc \ stars are shown on top of the isochrones. Our most important limits come from the luminosity ratios and the total photometry for the binary. While metallicity affects the luminosity of the stars in the last stages of core hydrogen burning (in the sense that more metal poor models are more luminous), the amount of convective core overshooting appears to be an important factor in the models shown here. PARSEC models have the largest amount of core overshooting ($\sim0.25H_P$ for $M > 1.5\msun$) among the models examined here, and the predicted luminosities of the \epc \ components appear to be too high to reproduce \epc. (Alternately, if the luminosities are forced to match the observations, the primary star is forced to be too red, and the color of the binary can't be matched.) The MIST and BaSTI-IAC evolution tracks for the masses of the \epc \ components  pass through the observed magnitude levels toward the end of core hydrogen burning, which occur near the red kinks in each track. The redward end of the kink approximately matches the CMD positions of the bright single main sequence stars in Praesepe as well. As shown in Figure \ref{cmd_decomp}, MIST models match the inferred $G$ magnitudes of both components to within the uncertainties in the range $632-641$ Myr if the masses are as measured and the cluster metallicity roughly matches the value from the literature ([Fe/H]$=+0.14$). When the model photometry values are combined, the measured {\it Gaia} photometry for the system is also matched.

Uncertainties on the measured masses and cluster metallicity extend the range of acceptable ages. Lower primary star masses and higher metallicity models require older ages.
The primary star mass uncertainty results in about 15 Myr of age uncertainty.
At the ends of the mass and metallicity uncertainty ranges, color changes in the primary star lead to unsatisfactory matches to the observed color of the binary. This is a stronger effect for the metallicity, with [Fe/H] values more than about $\pm0.03$ dex (approximately $0.5\sigma$) away from the spectroscopic value producing stellar colors that are too blue or red. A $0.5\sigma$ uncertainty in metallicity results in an age uncertainty of about 10 Myr. Adding these uncertainties in quadrature, we therefore quote an age uncertainty of 19 Myr. Perhaps more importantly than the statistical uncertainty, we believe the observational limits reduce systematic errors in the age by ruling out some isochrone sets (and the model physics they utilize) as unable to reproduce the characteristics of \epc. Smaller changes to the amount of convective core overshooting could still produce model stars consistent with the characteristics of the components of \epc \ at different ages (with larger amounts of overshooting resulting in younger ages). We also note that the initial He abundance of these metal-rich Praesepe stars carries some uncertainties that affect the detailed comparisons with models, including the color of the main sequence and the main sequence lifetime of turnoff stars. (For a discussion of this issue in the even more metal-rich cluster NGC 6791, see \citet{brogaard1,brogaard2}.) Evaluation of these contributions to the age uncertainty is beyond the scope of this study, however.

A MIST isochrone of approximately 635 Myr
and the spectroscopically-measured cluster metallicity can simultaneously match the blue edge of the distribution of bright main sequence stars, the red extent of the red kink at the cluster turnoff, and the magnitudes of the components of \epc \ at their measured masses.

\begin{figure}
\plotone{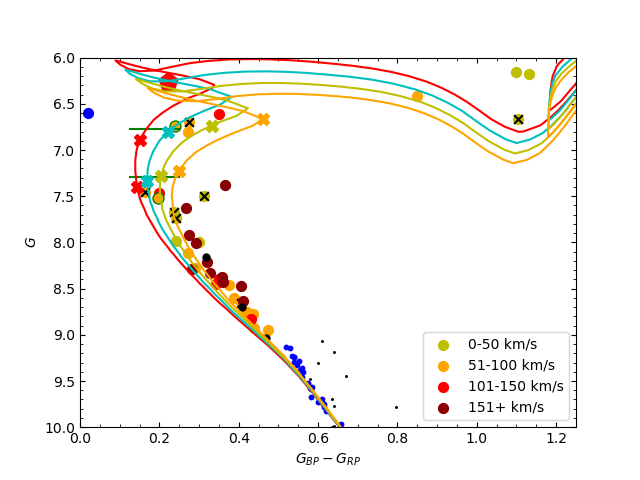}
\caption{PARSEC isochrones \citep{2012MNRAS.427..127B} for ages 600 (red), 650 (cyan), 700 (yellow), and 750 (orange) Myr. In the graph, distance modulus $(m-M)_0= 6.33$, iron abundance [Fe/H]$ = +0.16$, and reddening $E(G_{BP}-G_{RP})= 0.073$ were used. $\times$ markers indicate binary systems.The colored $\times$ markers indicate where our calculated masses in table \ref{tab:chiparams} land on the isochrone.} \label{parsec}
\label{parsec}
\end{figure}

While it is gratifying that models reproduce several important features of the CMD, a careful examination raises some questions. 
The isochrone does not match the photometry of the star HD 72779 in the Hertzsprung gap. (See Appendix \ref{app_pra} for evidence that the star is single, and representative of where the subgiant portion of the isochrones should be.) This may indicate some mismatch in the hydrogen abundance profile near the core at hydrogen exhaustion.
The distribution of stars in the range $6.5 < G < 7.5$ is also difficult to interpret with conventional stellar models (Fig. \ref{mist}-\ref{parsec}). The stars at $G \approx 6.7$ and 7.5 are consistent with a single age of around 635 Myr, with the first group of stars at the turnoff and the second group near the reddest point in the isochrone's kink (Fig. \ref{mist}). According to the evolutionary timescales, both parts of the isochrone should be fairly well populated. However, the range $6.8 < G < 7.4$ should also be occupied as it should still contain stars burning core hydrogen, and we find no stars in this range. Additionally, most of the bright main sequence stars in Praesepe ($7.5 < G < 10$; $1.2 < M_G < 3.7$) reside brighter or redder than the model isochrones, although the amount of disagreement varies between model sets and can be affected by chemical composition (metallicity or He abundance). In addition, there appears to be a jag in Praesepe main sequence at $G\approx9.1$ ($M_G \approx 2.8$), and slightly fainter stars seem to deviate noticeably from the models. An examination of the Hyades (below) does not give us clear evidence for a similar feature. The Praesepe feature appears to be about a magnitude too bright to be related to the Kraft break \citep{kraft} in stellar rotation rates.

\begin{figure}
\plotone{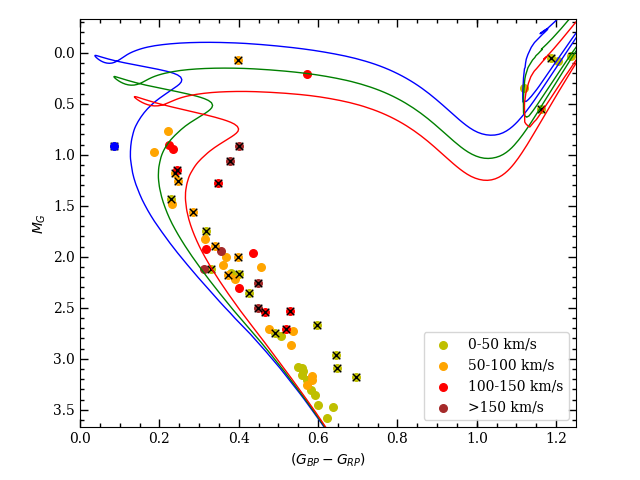}
\caption{Hyades members in an color-absolute magnitude diagram using {\it Gaia} EDR3 photometry and parallaxes. 
Color coding for individual stars is based on measured rotational velocity $v_{rot} \sin i$, and $\times$ symbols indicate binary systems. The blue point is the blue straggler $\delta^3$ Tau.
MIST isochrones \citep{2016ApJ...823..102C} for [Fe/H]$=+0.14$ and zero reddening are shown, with ages 635 (blue), 740 (green), and 850 (red) Myr shown.
\label{hyades}}
\end{figure}

The Hyades cluster provides a useful comparison, as it has nearly identical age and chemical composition, and a similar richness of stars. Bright Hyades members are plotted in Figure \ref{hyades}, with the plot ranges chosen to cover the ranges in the Praesepe plots. There are no stars found at the luminosity level of the brighter group ($G \approx 6.7$ in Praesepe, $M_G \approx 0.4$), and a handful of stars are scattered above the bright end of the main sequence in Praesepe ($G \approx 7.5$, $M_G \approx 1.2$). Most of these bright stars have shown some evidence of binarity in the literature, but we discuss in Appendix \ref{app_hy} why we believe the photometry for most of the stars with $M_G < 1.6$ and $(G_{BP}-G_{RP})< 0.3$ is probably representative of single stars. Similarly to the situation for Praesepe, there appear to be almost no stars present in what should be the last stages of core hydrogen burning, between the turnoff and the red kink.

In summary, the component stars of \epc, and the two groups of late main sequence stars push for an age around $637\pm19$ Myr,
but the single star in the Hertzsprung gap and the gap in the star distribution may imply a statistical fluke or possibly an error in the stellar model physics or chemical composition. In addition, the PARSEC isochrones generally predict too high a luminosity for a star of \epc \ A's mass in the cluster, while the MIST and BaSTI-IAC isochrones reproduce the luminosity better. As a result, we believe our analysis puts limits on the amount of convective core overshooting used in the isochrones, and therefore reduces systematic uncertainties in the cluster age. We encourage work to analyze additional stars in Praesepe in order to identify the cause of these inconsistencies.

\subsection{White Dwarf Progenitor Masses}

A primary result of the measurements of \epc \ is the precise measurement of the mass of a star near the end of core hydrogen burning. This significantly reduces the uncertainties in the predicted masses of more evolved stars, and more cleanly separates those predicted masses from uncertainties surrounding the cluster age. While there are still some uncertainties surrounding subsequent evolutionary phases, this can improve the determination of white dwarf progenitor masses in the cluster.

For a single-age cluster, Praesepe contains a sizable sample of white dwarfs with measured masses covering a fairly large range from about 0.7 to $0.9 \msun$ \citep{cummingswd,salaris}. These white dwarfs are in a transitional range between the masses of common field white dwarfs with CO compositions and massive white dwarfs above $1 \msun$ presumed to have ONe compositions. The progenitor stars are believed to be ones that ignite helium burning in a non-degenerate core, but do not experience a second convective dredge-up in their asymptotic giant stage. The second dredge-up limits the growth of the hydrogen-depleted cores in more massive stars, and results in slower growth of white dwarf masses with increasing initial star mass. More precise determinations of the initial masses for Praesepe's white dwarfs can produce a strong lower mass limit for stars that undergo the second dredge up, as well as a stronger test of our understanding of the initial-final mass relation (IFMR) in this transition range.

To determine initial masses for these white dwarfs, we started with isochrones for which a star of \epc \ A's mass passed through the range of $G$ magnitudes we limited the star to earlier in this paper. We then subtracted cooling times for the Praesepe WDs from \citet{cummingswd} from the isochrone ages, and calculated isochrones for those younger ages. The progenitor masses for the white dwarfs are then the initial masses of stars ending the asymptotic giant branch at that earlier time. Although the initial masses derived in this way are still somewhat subject to uncertainties in evolutionary phases after the main sequence (most notably during the relatively long core helium burning phase), the progenitor masses are more explicitly tied to masses measured for present-day stars. We based our calculations on PARSEC models exclusively because of the difficulties other isochrone sets had in reproducing the CMD position of \epc \ A, and these difficulties probably result from differences in convective core overshooting algorithms. Based on the new age estimation of
$637\pm19$ Myr, the new initial masses are given in Table \ref{ifmrtab}.

\begin{deluxetable}{cccc}
\tablecaption{Initial Mass Determinations for Praesepe WDs}
\tablehead{\colhead{ID} & \colhead{$M_{WD} / \msun$\tablenotemark{a}} & \colhead{$t_{cool}$ (Myr)\tablenotemark{a}} &  \colhead{$M_i / \msun$}}
\startdata
WD0833+194 & $0.813\pm0.027$ & $364^{+33}_{-30}$ & $3.71^{+0.19}_{-0.14}$\\
WD0836+199 & $0.830\pm0.028$ & $352^{+34}_{-31}$ & $3.66^{+0.19}_{-0.11}$\\
WD0837+189 & $0.870\pm0.029$ & $402^{+42}_{-39}$ & $3.92^{+0.29}_{-0.23}$\\
WD0837+199 & $0.757\pm0.025$ & $190^{+16}_{-15}$ & $3.16^{+0.06}_{-0.06}$\\
WD0840+190 & $0.895\pm0.030$ & $425^{+48}_{-44}$ & $4.07^{+0.43}_{-0.30}$\\
WD0840+200 & $0.752\pm0.027$ & $233^{+19}_{-18}$ & $3.27^{+0.07}_{-0.07}$\\
WD0843+184 & $0.898\pm0.028$ & $423^{+38}_{-41}$ & $4.05^{+0.32}_{-0.28}$\\
\enddata
\tablenotetext{a}{\citet{cummingswd}.}
\label{ifmrtab}
\end{deluxetable}

In their examination of the IFMR, \citet{cummingswd} derived ages for Praesepe of $705\pm25$ Myr and $685\pm25$ Myr from non-rotating PARSEC and MIST isochrones, respectively for [Fe/H]$=+0.15$ but zero reddening. \citet{salaris} used the age derived by \citet{gaiahrs}: $708^{+140}_{-90}$ Myr for [Fe/H]$=+0.16$ and reddening $E(B-V)=0.027$. Even with high-quality {\it Gaia} photometry, there is definite ambiguity in fitting the CMD with isochrones due to the small number of bright stars and potentially confounding factors like binarity and high rotation rates.
Our lower age results in higher initial masses for the cluster white dwarfs by $0.14-0.31\msun$ relative to the MIST results

from \citeauthor{cummingswd}
For all but the youngest white dwarfs, the cooling time uncertainty is the largest contributor to the uncertainty in the initial masses. \citet{salaris} derived cooling times from fitting tracks to Gaia DR2 photometry of the white dwarfs, and found values that were in most cases smaller (by $\sim80-160$ Myr) than those of \citet{cummingswd}. \citeauthor{cummingswd} derived cooling times from spectroscopic measurements of $\teff$ and $\log g$. As a result, it should be remembered that systematic uncertainties in the cooling times could have a larger effect on the initial masses than our revision of the cluster age.

We believe that our determinations represent an improvement in how realistic the statistical uncertainties are (because \citeauthor{cummingswd} appear to have only included cooling time uncertainties in their quotations of initial mass uncertainties) and in reducing systematic uncertainties (because the measured star masses allow us to identify isochrones that match masses as well as CMD properties).

\section{Conclusion}\label{sec:conclusion}

Our overall goal was to determine the age of the open star cluster Praesepe using the binary star system \epc. Using radial velocities and interferometric measurements, we have been able to calculate the primary star’s mass as $2.420\pm0.008 M_\odot$ and the secondary star’s mass as $2.226\pm0.004 M_\odot$, as well as put observational limits on the brightness of the two stars. 
With this information, we find that MIST models with the spectroscopic cluster metallicity are able to successfully reproduce the inferred CMD positions of the stars of \epc \ at their masses and the position of the pre-core hydrogen exhaustion kink. While this gives evidence in favor of the commonly used amount of convective core overshooting in the MIST models, there are other features (the luminosity of a cluster subgiant, and an unusual distribution of stars brighter than the cluster turnoff) that may be indicating issues with model physics that should be investigated further.
The MIST models indicate an age of $637\pm19$ Myr for the cluster.

The direct measurement of masses for stars nearing core hydrogen exhaustion also allows us to re-examine the likely progenitor masses for the cluster white dwarfs. Our measurements imply an upward shift to the initial masses for white dwarfs in the $0.7-0.9 \msun$ range, although systematic uncertainties in cooling ages need to be addressed and could have a substantial effect.

\acknowledgements
We gratefully acknowledge the assistance of Daryl W. Willmarth in locating and providing 
Kitt Peak spectra of \epc \ from more than 25 years ago, and support from the National Science
Foundation under grant AAG 1817217 to E.L.S.
This research made use of observations from the SIMBAD database,
operated at CDS, Strasbourg, France; the WEBDA database, operated
at the Institute for Astronomy of the University of Vienna; data
products from the AAVSO Photometric All Sky Survey (APASS), funded by
the Robert Martin Ayers Sciences Fund and the National Science
Foundation (grant AST-1412587). We use observations made with the Nordic Optical Telescope, owned in collaboration by the University of Turku and Aarhus University, and operated jointly by Aarhus University, the University of Turku and the University of Oslo, representing Denmark, Finland and Norway, the University of Iceland and Stockholm University at the Observatorio del Roque de los Muchachos, La Palma, Spain, of the Instituto de Astrofisica de Canarias.

This work is based upon observations obtained with the Georgia State University Center for High Angular Resolution Astronomy Array at Mount Wilson Observatory.  The CHARA Array is supported by the National Science Foundation under Grant No. AST-1636624 and AST-2034336.  Institutional support has been provided from the GSU College of Arts and Sciences and the GSU Office of the Vice President for Research and Economic Development. Time at the CHARA Array was granted through the NOIRLab community access program (NOAO PropIDs: 2017A-0273, 2018A-0286, 2018B-0218, 2020A-0379; NOIRLab PropID: 2021A-0276; PI: E. Sandquist). MIRC-X received funding from the European Research Council (ERC) under the European Union's Horizon 2020 research and innovation programme (Grant No. 639889). J.D.M. acknowledges funding for the development of MIRC-X (NASA-XRP NNX16AD43G, NSF-ATI 1506540, NSF-AST 1909165). S.K. acknowledges support from an ERC Consolidator Grant (Grant Agreement No.\ 101003096). A.L. received support from a Science Technology and
Facilities Council (STFC) studentship (No. 630008203).  LRB acknowledge support by MIUR under PRIN program $\#$2017Z2HSMF.

\facilities{CHARA (CLIMB, MIRC-X), KPNO:CFT, NOT (FIES), OHP (ELODIE)}
\software{BF-rvplotter (https://github.com/mrawls/BF-rvplotter), FIESTool}

\begin{deluxetable}{l|cc|cl}
\tablecaption{Literature Radial Velocity Measurements for \epc \ A \label{tab:1}}
\tablehead{\colhead{mJD\tablenotemark{a}} &  \colhead{$v_{r,A}$} & \colhead{$\sigma_v$} & \colhead{Phase $\phi$} & \colhead{Source}\\
& \multicolumn{2}{c}{\kms} & }
\startdata
19849.3\tablenotemark{b} & 17 & 5 & 0.971 & Potsdam\\
19852.3\tablenotemark{b} & 73 & 5 & 0.057 & Potsdam\\
23451.941 &  32.6 & 9.6  & 0.489 & MWO\\
23534.651 & $-21.3$ & 2.5  & 0.843 & MWO \\
23539.721 &  19.3 & 16.9 & 0.987 & MWO\\
23563.646 &  14.4 & 3.5  & 0.668 & MWO\\
23573.664 & $-16.4$ & 12.4  & 0.953 & MWO\\
48309.755 & $-28.5$ & 3.5  & 0.853 & KPNO \\
48312.724 & $-17.9$ & 3.5  & 0.938 & KPNO\\
48330.863 &  44.2 & 3.5  & 0.454 & KPNO\\
48580.917 &  27.5 & 3.5  & 0.569 & KPNO \\
48673.845 &  78.3 & 3.5  & 0.214 & KPNO\\
48676.736 &  65.3 & 3.5  & 0.296 & KPNO\\
48694.761 & $-17.6$ & 3.5  & 0.809 & KPNO\\
48695.800 & $-22.4$ & 3.5  & 0.839 & KPNO\\
48969.909 &  16.9 & 3.5  & 0.639 & KPNO\\
49107.740 &  32.4 & 7    & 0.561 & KPNO\\
\enddata
\tablenotetext{a}{mJD = HJD $- 2400000$}
\tablenotetext{b}{Estimated, as time of observation is not available.}
\end{deluxetable}

\begin{deluxetable}{l|rc|rc|cl}
\tablecolumns{6}
\tablecaption{Radial Velocity Measurements for \epc ~ A and B\label{tab2}}
\tablehead{ \colhead{mJD\tablenotemark{a}} &  \colhead{$v_{r,A}$} & \colhead{$\sigma_v$} &\colhead{$v_{r,B}$} & \colhead{$\sigma_v$} & \colhead{Phase $\phi$} & \colhead{Source} \\ & \multicolumn{2}{c}{\kms} & \multicolumn{2}{c}{\kms} &}
\startdata
22384.753 &  85.4 &  3  & $-20.1$ & 3 & 0.121 & DAO\\
48613.92562 &  36.5 & 1.5 & 24.7 & 2.8 & 0.509 & KPNO\\
48638.82381 &  79.9 & 1.1 & $-17.4$ & 2.7 & 0.217 & KPNO\\
48639.93300 &  76.3 & 1.1 & $-12.5$ & 2.7 & 0.249 & KPNO\\
48674.83253 &  73.4 & 1.0 & $-18.4$ & 2.9 & 0.242 & KPNO\\
48696.76600 & $-27.9$ & 0.6 & 95.6 & 2.2 & 0.866 & KPNO\\
48967.99971 &  25.5 & 1.3 & 36.4 & 3.0 & 0.584 & KPNO\\
49019.89793 &  70.1 & 1.1 & $-9.1$ & 2.9 & 0.061 & KPNO\\
49021.87432 &  84.6 & 0.9 & $-19.9$ & 2.5 & 0.117 & KPNO\\
49021.93111 &  85.3 & 0.9 & $-19.7$ & 2.4 & 0.119 & KPNO\\
49021.98021 &  85.6 & 1.0 & $-19.5$ & 2.3 & 0.120 & KPNO\\
49030.84515 &  58.7 & 1.3 & 6.3 & 3.7 & 0.373 & KPNO\\
49080.62158 & $-13.8$ & 1.1 & 83.6 & 2.8 & 0.789 & KPNO\\
49081.70621 & $-19.2$ & 1.0 & 90.5 & 2.6 & 0.820 & KPNO\\
49082.71360 & $-23.8$ & 1.0 & 94.9 & 2.8 & 0.849 & KPNO\\
49094.69748 &  83.5 & 1.0 & $-17.8$ & 2.3 & 0.190 & KPNO\\
49095.69961 &  79.4 & 1.0 & $-17.1$ & 2.7 & 0.218 & KPNO\\
49096.69919 &  76.3 & 1.1 & $-12.3$ & 3.0 & 0.247 & KPNO\\
49097.71513 &  72.9 & 1.0 & $-4.7$ & 3.0 & 0.276 & KPNO\\
49106.71287 &  33.8 & 1.4 & 35.3 & 2.7 & 0.532 & KPNO\\
49344.96554 &  67.7 & 1.2 & $-3.0$ & 3.1 & 0.311 & KPNO\\
49345.98097 &  64.2 & 1.3 & 2.8 & 3.2 & 0.340 & KPNO\\
49381.90408 &  60.6 & 0.9 & 9.0 & 3.6 & 0.363 & KPNO\\
49382.80479 &  56.7 & 1.4 & 10.4 & 3.5 & 0.388 & KPNO\\
49383.77115 &  53.2 & 1.5 & 14.6 & 4.1 & 0.416 & KPNO\\
49384.87950 &  46.3 & 1.4 & 25.2 & 3.1 & 0.447 & KPNO\\
50800.02879 &  1.1 & 1.3 & 62.4 & 3.7 & 0.717 & KPNO\\
53009.47942 &  23.9 & 0.1 & 45.6 & 2.2 & 0.590 & OHP\\
53014.62148 &  $-1.8$ & 0.1 & 75.0 & 1.9 & 0.736 & OHP\\
58960.37513 & $-19.8$ & 0.1 & 94.3 & 1.3 & 0.931 & NOT\\
58960.37905 & $-19.8$ & 0.1 & 94.1 & 1.3 & 0.931 & NOT\\
58968.40424 &  86.1 & 0.1 & $-20.4$ & 1.5 & 0.160 & NOT\\
58968.40826 &  86.0 & 0.1 & $-20.5$ & 1.5 & 0.160 & NOT\\
\enddata
\tablenotetext{a}{mJD = HJD $- 2400000$}
\tablenotetext{b}{The normalized digitized spectra for KPNO and NOT (30 spectra) are available as the Data behind the Figure.}
\end{deluxetable}



\clearpage
\bibliography{references}{}
\bibliographystyle{aasjournal}
\nocite{*}

\appendix

\section{Bright Main Sequence and Subgiant Stars in Praesepe and the Hyades}

In order to make the cleanest CMD comparison between the two clusters, we need to understand whether the photometry of the most age-sensitve stars is affected by binarity. Most of the bright stars have have been claimed to be close binaries, but in many cases the evidence is rather marginal. We discuss the current evidence below, proceeding from most to least luminous stars. 

\subsection{Praesepe Members\label{app_pra}}

{\it 35 Cnc / HD 72779 / VL 133:} This star is a fairly rapidly rotating subgiant star in the red half of the Hertzsprung gap. Radial velocity monitoring by \citet{aandw} indicate a constant radial velocity, while lunar occultation observations \citep{blow,eblo} and speckle interferometry \citep{speckle} have not shown evidence of a companion.


{\it 38 Cnc / BT Cnc / HD 73575 / KW 204:} \citet{aandw} report that this star has no velocity variability, but does show line profile variations. This may relate to its $\delta$ Sct variability. \citet{breger} finds light variations with full amplitude of about 0.06 mag. No companion was detected during lunar occultations \citep{occult1,occult2} or speckle interferometry \citep{speckle}.

{\it HD 73210 / KW 50:} \citet{aandw} derive a double-line spectroscopic orbit with a 12 day period for this star. The detection of both stars implies substantial light blending, so that the CMD position will not be representative of single stars. However, the binary has a fairly extreme mass ratio ($q \approx 0.35$) that means that the CMD position of the brighter star does not reside too far from that of the binary.
Lunar occultation \citep{blow} and speckle interferometry \citep{speckle} did not reveal a companion. 

{\it HD 73712 / KW 284:} \citet{aandw} observe a sharp-lined pair of stars in spectra, superimposed on a third component with broad lines. They derive a spectroscopic orbit with a 48 day period for the sharp-lined pair with a mass ratio  fairly close to one. Needless to say, the CMD position does not adequately represent any of the component stars, and all are likely to be near the main sequence. 
Speckle interferometry \citep{speckle,mca89,hart94,hart97,hart00,toko} and lunar occultation \citep{occult2} probably show orbital motion of the third star, approximately 1.6 mag fainter with about $0\farcs1$ angular separation.

{\it EP Cnc / HD 73819 / KW 348:} Similar to 35 Cnc, \citet{aandw} report that this star has no velocity variability but rapid line profile variations. This may also relate to its $\delta$ Sct variability. The amplitude of the stars's photometric variations is relatively small ($\sim0.01$ mag; \citealt{breger}), however. No companion was detected during lunar occultations \citep{occult1,occult2}.

{\it 42 Cnc / HD 73785 / KW 328:} \citet{aandw} report no velocity variation in their extensive observations, in contradiction to an earlier report by \citet{rm98} that it was an SB2 based on older observations.
No companion was detected during lunar occultations \citep{occult1,occult2}.

{\it HD 73449 / KW 150:} This star has long been known to rotate at greater than 200 \kms, and it is considerably redder than other cluster stars at the same brightness. A lunar occultation \citep{occult1} and speckle interferometry \citep{speckle} did not reveal evidence of a companion star. \citet{rm98} identified it as an SB1 on the basis of old radial velocity measurements, but \citet{aandw} found it to have constant radial velocity. The star has been identified as photometrically variable in the older literature, but recent studies have used it as a constant photometric comparison star for $\delta$ Sct studies \citep{hernandez,pena}.

{\it HD 72942 / KW 534:} HD 72942 is a known single-lined spectroscopic binary with an orbital period of 131.96 d \citep{aandw}, indicating a lack of interaction between the stars. The fact that the binary's color is bluer than any main sequence star also implies that the primary star must be one of the bluest stars in the cluster. The star has an interesting hybrid surface composition between normal A stars and Am stars \citep{2007AA...476..911F}. In addition, the primary star is rotating at a rate that puts it toward the high end of the range for the cluster's Am stars \citep{fossati08}. These items leave a somewhat ambiguous picture of the star's nature, but a blue straggler history cannot be ruled out. A lunar occultation \citep{occult1} and speckle interferometry observations \citep{speckle} were negative for a companion. We conclude that this star probably should not be considered a representative of single cluster stars.

{\it HD 72846: } This star had a composition study by \citet{fossati08}, who found it to be consistent with normal main sequence abundances in the cluster. No evidence of binarity for this star has yet been found in speckle interferometry observations \citep{speckle}, and a small number of radial velocity measurements are consistent with a constant value \citep{abt}.

{\it HD 73711 / KW 276:} This star has abundances consistent with chemically-peculiar Am stars \citep{2007AA...476..911F,bc98}, and as with many of those, there are tentative signs of binarity in spectra. \citet{bc98} identified the star as SB2 based on line profiles, while \citet{rm98} identified it as SB1 based on variation in older radial velocities. No velocity variations were found in the observations by \citet{aandw}, although they saw evidence of asymmetries in the line profiles. Lunar occultation \citep{occult1} and speckle interferometry \citep{speckle} did not reveal a companion.

\subsection{Hyades Members\label{app_hy}}

{\it $\delta$ Cas / 37 Cas / HD 8538:} This star is apart from the main body of the cluster, and was recently identified as a member of the preceding tidal tail of the cluster \citep{roser_tails}. Its position in the Hertzsprung gap makes it interesting for isochrone fitting. The star does not appear to be velocity variable \citep{abt65}. Although it is identified as an eclipsing variable in SIMBAD, there is no strong evidence that it is photometrically variable \citep{GCVS}. To date, the evidence is that it is a single star, although there is not high-resolution imaging or interferometry in the literature.

{\it $\theta^2$ Tau / 78 Tau / HD 28319 / vB 72:} This is a very well-known spectroscopic and interferometrically-resolved binary. The most recent analyses of the binary are in \citet{torres11} and \citet{armstrong}. Although the photometry of the two components can potentially be disentangled, this is beyond the scope of this work.

{\it $\kappa^1$ Tau / 65 Tau / HD 27934 / vB 54:} This star is identified as probably velocity constant in \citet{abt65}. The only reference to binarity is a report from a visual observation of a lunar occultation \citep{dunham_lo1} that reported the minimum resolvable separation ($0\farcs1$) and equal brightness components. Attempts at speckle interferometry detections have been negative \citep{patience,mason_hy,handm84}. We conclude that there is not significant evidence of binarity for this star.

{\it $\delta^3$ Tau A / 68 Tau / HD 27962 / vB 56:} This star is identified as a cluster blue straggler based on its very blue color, as well as an Am star, and so we do not consider it in discussing the cluster age. It is a velocity variable \citep{borgniet}, but there have only been negative detections of a close companion in lunar occultations \citep{trunkovsky} and speckle interferometry \citep{patience,mason_hy,handm84}.

{\it $\iota$ Tau / 102 Tau / HD 32301 / vB 129:} 102 Tau was shown to have constant radial velocity in \citet{abt65}. Speckle interferometry observations have not revealed a close companion \citep{handm84,mason_hy,patience}. There is a report of a lunar occultation in 1956 that indicated two equal-brightness stars at $0\farcs4$ separation \citep{bsc}, although this seems to contradict the negative speckle results. At present, the evidence for a companion star with significant brightness is minimal.

{\it c Tau / 90 Tau / HD 29388 / vB 104:} We ignore the wide common proper motion components in this discussion. A faint companion was detected interferometrically by \citet{pionier} at a separation of 11 mas, contributing only about 3\% in $H$ band. 90 Tau was reported early on as a spectroscopic binary \citep{barrett}, but is labeled as a "constant:" velocity star in \citet{abt65}. Based on the evidence, we expect the interferometric companion to have little effect on the system photometry in the optical, but it is unclear whether there is a closer companion affecting the velocities.

{\it $\sigma^2$ Tau / 92 Tau / HD 29488 / vB 108:} 92 Tau was identified by \citet{lee} as a double-lined spectroscopic binary, but later shown by \citet{abt65} to have constant velocity. \citet{borgniet} identify low-level velocity variations ($\sim1$ km s$^{-1}$) from more recent data covering 1370 days, and \citet{lagrange} identify it as velocity variable but not as a binary candidate. Speckle interferometric observations have not detected a companion \citep{patience,mason_hy}. We conclude that there may be a companion to the star, but that it is likely to be faint based on the lack of spectroscopic detections.

{\it $\lambda$ Gem / 54 Gem / HD 56537:} This is a star identified in the trailing tail of the Hyades \citep{roser_tails}. It was identified early as a spectroscopic binary with a period of $4-5$ days by \citet{hnatek}, and again by \citet{frost}, with other indications of velocity variations in \citet{abt}. \citet{lagrange} did not identify it as a binary candidate from more recent radial velocity measurements, nor did \citet{becker}. While there is a visual companion at $10\arcsec$ separation, a closer companion has been resolved by lunar occultation: initially by \citet{dunham_lo} and then by \citet{richi4} following several negative results. \citeauthor{dunham_lo} initially identified the closer companion as visually fainter by 1 mag, but \citeauthor{richi4} found it fainter in $K$ by about 3.4 mag. Both studies found separations of tens of milli-arcseconds. The companion was not detected in other ground-based interferometer measurements \citep{pionier}. Based on the evidence, we expect the companion to have a small effect on the system photometry.

{\it $\theta$ Cas / 33 Cas / HD 6961:} This star is in the preceding tidal tail of the cluster \citep{roser_tails}, and was identified as a long-period single-lined spectroscopic binary by \citet{abt65} and \citet{aandl}. More recently, \citet{borgniet} found no significant radial velocity variation over a 1417 d baseline. Speckle interferometry studies have not revealed a close companion \citep{handm84}. The star is used as a calibrator for interferometric studies, and there have been no reports of binarity. We conclude that there may be low-level contamination of the light of the primary star.

{\it HD 30210 / vB 112:} We will ignore two wide commmon proper motion companions in this discussion. This star was identified in the trailing tail of the cluster \citep{roser_tails}. It is also a known Am star, which generally have higher probabilities of binarity than single stars. \citet{aandl} identify it as a single-lined spectroscopic binary with uncertain orbital elements. Speckle observations \citep{mason_hy} gave negative results for a companion. Based on the potential color effects of being an Am star and the chance that the spectroscopic companion might affect the photometry, we leave it out of consideration for age purposes.

{\it $\delta^2$ Tau / 64 Tau / HD 27819 / vB 47:} This star was identified as a likely spectroscopic binary by \citet{lee}, but was identified as velocity constant in \citet{abt65} with a chance of slow variation. \citet{borgniet} find evidence of low amplitude velocity variations over a 1399 day span, but inconsistent with the earlier \citeauthor{lee} measurements. A lunar occultation observation \citep{africano} and speckle interferometry \citep{handm84,mason_hy} were negative for a companion. We conclude that there may be low-level effects on the photometry of the primary star.

{\it HD 28527 / vB 82:} This star is labelled as a $\delta$ Sct variable in SIMBAD, but it appears to be a constant star \citep{kp89}. It was identified as a possible binary by \citet{barrett}, but later labeled as velocity constant in \citet{abt65}. Multiple lunar occultation observations \citep{white79,radick,fekel,eande,radick82,richi4} have been negative for a companion, while \citet{peterson81} provide "strong, but not definitive" evidence of a companion with fairly large magnitude difference ($\sim3$ mag) at very small separation. We conclude that the weight of the evidence is against a bright, close companion.


\end{document}